# The effects of shear and near tip deformations on interface fracture of symmetric sandwich beams

Luca Barbieri\*, Roberta Massabò\*,^, Christian Berggreen\*\*

\* Department of Civil, Chemical and Environmental Engineering,
University of Genova, Via Montallegro 1, 16145, Genoa, Italy
^ Corresponding author: mailto:roberta.massabo@unige.it

\*\* Department of Mechanical Engineering, Technical University of Denmark,
Nils Koppels Allé Building 404, 2800 Kongens Lyngby, Denmark

**Abstract**

The effects of shear on energy release rate and mode mixity in a symmetric sandwich beam with isotropic layers and a debond crack at the face sheet/core interface are investigated through a semi-analytic approach based on two-dimensional elasticity and linear elastic fracture mechanics. Semi-analytic expressions are derived for the shear components of energy release rate and mode mixity phase angle which depend on four numerical coefficients derived through accurate finite element analyses. The expressions are combined with earlier results for three-layer configurations subjected to bending-moments and axial forces to obtain solutions for sandwich beams under general loading conditions and for an extensive range of geometrical and material properties. The results are applicable to laboratory specimens used for the characterization of the fracture properties of sandwich composites for civil, marine, energy and aeronautical applications, provided the lengths of the crack and the ligament ahead of the crack tip are above minimum lengths. The physical and mechanical significance of the terms of the energy release rate which depend on the shear forces are explained using structural mechanics concepts and introducing crack tip root rotations to account for the main effects of the near tip deformations.

**1. Introduction**

Composite sandwich structures are widely used in marine, energy, aeronautical and civil engineering applications. Their main advantages over traditional metallic materials or monolithic composites are the high stiffness to weight and strength to weight ratios, which make them key enablers for present and future lightweight structures.

Sandwich structures, however, are highly susceptible to manufacturing flaws as well as in-service damages. One of such flaws and damages is debonds, which may arise at the face/core interfaces and will degrade the load carrying capacity and integrity of the sandwich structure, and may even grow catastrophically during both quasi-static and fatigue loading depending on their debond location and loading scenario in the structure.

In order to assess the criticality of debond flaws and damages, the fracture properties of the face/core interface have to be measured and used as input properties in fracture mechanical analysis models. Several mixed mode fracture mechanical characterization tests have been proposed in the literature and are currently used for fracture characterization of sandwich face/core interfaces (see review in [1]). The analysis of most of the proposed test specimens relies on approximate structural theories and/or numerical finite element analyses to define energy release rate and mode mixity phase angle. Knowing energy release rate and mode mixity angle allows the reduction of the fracture toughness and crack propagation rate vs. energy release rate from measured data for different crack tip mixed mode loading conditions.

Semi-analytic solutions based on 2D elasticity for sandwich beams subjected to generalized end forces are already available in the literature [2][3]. These solution, however, are limited to loading by pure bending and axial forces and are therefore applicable only to a limited number of fracture mechanics specimens, such as the Double Cantilever Beam specimen with Uneven Bending Moments , DCB-UBM [2][4]. However, most of the fracture specimens for sandwich composite systems are also subjected to shear, e.g. the Single Cantilever Beam (SCB) specimen, the Double Cantilever Beam (DCB) sandwich specimen, the Cracked Sandwich Beam (CSB) specimen, and the Mixed Mode Bending (MMB) sandwich specimen, and shear is expected to strongly modify the crack tip conditions and the fracture parameters, as shown for the SCB specimen in [5], as well as in bi-material systems with interface cracks and edge-cracked homogeneous layers [7][8].

The semi-analytic expressions derived in [2] [3] for energy release rate and mode mixity phase angle in symmetric and asymmetric sandwich specimens subjected to axial forces and bending moments were obtained following and extending the method originally formulated by Suo and Hutchinson for isotropic, orthotropic and bi-material elastic layers [6][12][13][11]. The energy release rate is derived in closed form and coincides with predictions based on classical structural theories which assume the delaminated beam arms to be rigidly clamped at the crack tip cross section. In this problem the energy release rate is not affected by the near tip deformations or by the work done in process zones ahead of the crack tip and depends only on the applied forces and the material and geometrical properties of the arms. The semi-analytic expressions of the mode mixity angle, on the other hand, require the numerical derivation of a single, load independent, dimensionless coefficient, which is typically chosen as the mode mixity angle associated to loading by pure axial forces.

The presence of shear forces substantially modify the response. Shear deformations along the beam arms, near tip deformations and nonlinear mechanisms at the crack tip affect both the energy release rate and the mode mixity angle and must be accounted for in the derivations. The effects of shear on the fracture parameters has been studied using LEFM concepts in [7] for isotropic bi-material layers and in [8] for edge-cracked orthotropic layers. Using two different approaches, which will be reviewed and applied later in the paper, semi-analytic expressions were derived in [7][8] for energy release rate and mode mixity phase angle. The expressions depend on four numerically derived coefficients which add to that required for axial/bending loading.

In this work we will review the formulation in [2][3][4] to define dimensionless expressions for energy release rate and mode mixity phase angle in a symmetric three-layer sandwich specimen subjected to bending moments and normal forces. We then derive semi-analytic expressions for energy release rate and mode mixity phase angle in beams subjected to crack tip shear forces. The derivation relies on dimensional analysis, two-dimensional elasticity and a displacement extrapolation technique, the CSDE method formulated in [9], which is implemented in a commercial finite element code and is necessary to define the four numerical dimensionless coefficients. The coefficients are presented in tabular form for a wide range of material and geometrical properties which describe both conventional composites and extreme composites with a very large mismatch of the elastic properties. The expressions are then combined with previous results [2] to obtain solutions applicable to beams subjected to arbitrary loading conditions. The physical and mechanical significance of the different terms of the energy release rate which depend on the shear forces are explained using structural mechanics concepts and introducing crack tip root rotations to describe the main effects of the near tip deformations on the fracture parameters. A DCB sandwich specimen is analyzed to further clarify these concepts. The accuracy of the semi-analytic solutions is verified through comparison with published results for bi-material and sandwich beams subjected to crack tip bending and shear and with accurate FE solutions for DCB sandwich specimens.

**2. Sandwich beam with a face/core debond**

**2.1 Assumptions, dimensionless groups and load decomposition**

The problem studied in this paper is schematized in Fig. 1. A symmetric sandwich beam with a traction-free interface crack, or face/core debond, is subjected to generalized end forces applied per unit width. The geometry is defined by the thickness of the upper and lower layers, $h_1$, and the thickness of the core or inner layer, $h_c$, and by the lengths of the crack and the intact portion ahead of the crack tip, $a$ and $c$, with $a, c \gg h_1, h_c$ to ensure that the stress distribution near the crack tip is unaffected by the actual distribution of the applied tractions and that the stresses from the crack tip do not interact with the end sections. Prescriptions on the minimum admissible lengths can be found in [2] for sandwiches subjected to axial forces and bending moments and will be given later in the paper for general loading cases. The layers are isotropic and linearly elastic with elastic constants which may differ, even substantially, between the inner and outer layers. Focus in this paper is on sandwich configurations, thus where the face sheets are stiffer than the core.

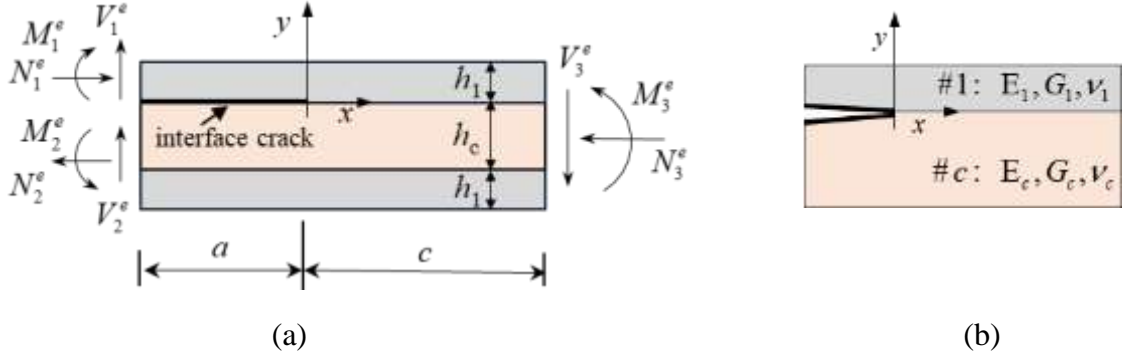

(a)                                                   (b)

Figure 1: (a) Symmetric sandwich beam with a face/core interface crack subjected to generalized end forces acting at the neutral axes of the respective cross sections. (b) Convention for the interface crack

Using the original notation in [11][12][6] the geometrical and material dimensionless groups which control the fracture behavior of the sandwich are:

$$\eta = h_1/h_c \, , \, \tilde{a} = a/h_c \, , \, \tilde{c} = c/h_c \tag{1}$$

$$\Sigma = \bar{E}_1/\bar{E}_c \text{ or } \alpha = \frac{\Sigma-1}{\Sigma+1} = \frac{\bar{E}_1 - \bar{E}_c}{\bar{E}_1 + \bar{E}_c},$$

$$\beta = \frac{G_1(\kappa_c - 1) - G_c(\kappa_1 - 1)}{G_1(\kappa_c + 1) + G_c(\kappa_1 + 1)} \text{ and } \varepsilon = \frac{1}{2\pi}\ln\left[\frac{(1-\beta)}{(1+\beta)}\right]$$

$$\left(\beta = \frac{1}{2}\frac{G_1(1-2\nu_c) - G_c(1-2\nu_1)}{G_1(1-\nu_c) + G_c(1-\nu_1)} \text{ for plane strain}\right)$$

where $\bar{E}_i = E_i$ for plane-stress and $\bar{E}_i = E_i/(1-\nu_i^2)$, for plane-strain, with $E_i$ and $\nu_i$ the Young's modulus and Poisson ratio of the layer $i$ and $G_i = E_i/(2+2\nu_i)$ the shear modulus; $\alpha$ and $\beta$ are the Dundur's parameters of the interface between the core and the upper face sheet, with $\kappa_i = 3 - 4\nu_i$ for plane strain and $\kappa_i = (3-\nu_i)/(1+\nu_i)$ for plane stress; $\varepsilon$ is the oscillatory index which characterizes the crack tip fields at the bi-material interfaces (see Section 2.2).

The Dundurs' parameter $\alpha$ is delimited by $|\alpha| \leq 1$, and assumed to be $0 \leq \alpha \leq 1$ in this paper to include only sandwich configurations where the face sheets are stiffer than the core, for $\Sigma > 1$, and homogeneous beams, for $\Sigma = 1$ and $\alpha = 0$. The admissible values of $\beta$, which correspond to non-negative Poisson's ratios, are defined by $-1 \leq \alpha - 4\beta \leq 1$, for plane strain, and $-1 \leq 3\alpha - 8\beta \leq 1$, for plane stress [20]; $\beta$ vanishes in a homogeneous layer or when both materials are incompressible, for $\nu_1 = \nu_c = 0.5$; and $\beta = (1-2\nu)\alpha/(2-2\nu)$ for identical Poisson ratios and plane strain, which particularize in $\beta = \alpha/4$ for $\nu_1 = \nu_c = 1/3$ ($\beta = \alpha/3$ for plane stress). As observed in [19] for typical material combinations in composite materials the parameter $\alpha$ is limited by $|\alpha| \leq 0.6$. However in most structural sandwich configurations, e.g. with foam, balsa and honeycomb cores, $\alpha$ is typically $\alpha \geq 0.8$ and approaches $\alpha = 1$ when the face material is extremely stiff compared to the core. In a sandwich with isotropic face sheets and core having $E_1 = 70$ GPa, $E_c = 7$ GPa, $\nu_1 = 0.3$ and $\nu_c = 0.32$, which could approximate a sandwich with aluminum face sheets and an aluminum foam core, $\alpha = 0.82$ (or $1/\Sigma = 0.12$) and $\beta = 0.23$; if $E_1 = 40$ GPa, $E_c = 0.2$ GPa, $\nu_1 = \nu_c = 0.3$, which could approximate a three-layer with glass fiber face sheets and a polymer foam core, then $\alpha = 0.99$ (or $1/\Sigma = 0.006$) and $\beta = 0.28$.

The generalized end forces in Fig. 1 are the axial forces $N_i^e$, bending moments, $M_i^e$, and shear forces, $V_i^e$, with $i = 1, 2, 3$ the index related to the three end sections ($i = 1$ debonded arm; $i = 2$ base arm; $i = 3$ base arm). The nine generalized components can be defined in terms of only six independent quantities by imposing global equilibrium. The axial forces are assumed to be applied at the neutral axes of the respective cross sections so that no coupling arise between extensional and bending response. The neutral axes of the debonded ($i = 1$) and base ($i = 3$) layers coincide with their geometrical centroidal axes, due to symmetry; the neutral axis of the composite substrate ($i = 2$) layer is defined in this paper by the distance from the centroidal axis of the core layer, $e_s$ Fig. 2a, and is given in Eq. (35) of the Appendix A1 as function of the dimensionless geometrical and material groups.

Due to the presence of the shear forces, $V_i^e$, with $i = 1, 2, 3$, the energy release rate and mode mixity will depend on the lengths of the crack and the ligament ahead of the crack, $a$ and $c$. General solutions applicable to all cases can be defined in terms of force and moment resultants acting at the crack tip cross sections, $N_i = N_i^e$ and $V_i = V_i^e$ for i=1,..3, $M_1 = M_1^e + V_1^e a$, $M_2 = M_2^e - V_2^e a$,

$M_3 = M_3^e - V_3^e c$. These generalized forces are then separated into two load systems: normal forces and bending moments, Fig. 2(b), and shear forces, Fig. 2(c).

Following the methodology in [6], which is based on the observation that when the transverse stresses are zero (Fig. 2b) the interface crack does not disturb the stress distribution, the load system in Fig. 2(b) is conveniently decomposed into the elementary systems: (b1) pure bending moments, $M$, and normal forces, $P$, acting on the two delaminated arms plus the required compensating moment $M_*$; (b2) pure bending moments and normal forces, $N_3$ and $M_3$, acting on the intact element. The sub-system (b2) is not involved in the calculation of the fracture parameters and in beams subjected to axial forces and bending moments only the response is then controlled by only two elementary load systems, $P$ and $M$.

Following the methodology in [7] the load system in Fig. 2c is also decomposed into the elementary systems: (c1) double-shear force, $V_D$, acting on the two arms and (c2) single-shear force, $V_S$, acting on the upper arm plus the compensating shear force acting on the base section. The elementary systems are related to the generalized crack tip forces, Figs. 2b and 2c, by:

$$P = N_1 - N_3 C_1 - \frac{M_3}{h_1} C_2 \qquad (2)$$

$$M = M_1 - M_3 C_3$$

$$M_* = M + P h_1 (\frac{1}{2} + \frac{1}{2\eta} + \tilde{e}_S)$$

$$V_D = -V_2$$

$$V_S = V_3$$

with:

$$C_1 = \frac{\Sigma \eta}{1 + 2\Sigma \eta} \qquad (3)$$

$$C_2 = \frac{1}{\tilde{D}_b} \left( \frac{1}{2} + \frac{1}{2\eta} \right)$$

$$C_3 = \frac{1}{12 \tilde{D}_b}$$

where $\tilde{e}_s$ and the dimensionless bending stiffness of the base cross section, $\tilde{D}_b$, are given in the Appendix 1, Eqs. (35),(38) [10]. The procedure applied to define the quantities in Eqs. (3), which is based on the superposition in Fig. 2b, was originally proposed in [6] for a bi-layer and extended to a symmetric sandwich in [2][1].

## 2.2 Local fields at the tip of the interface crack

The stress field and the relative crack displacements in the singularity dominated zone at the tip of the interface crack in Figs. 1,2a can be defined in terms of a complex stress intensity factor $K = K_1 + iK_2$ and an oscillatory index $\varepsilon$, Eq. (1), [14][15]. With reference to a Cartesian system $x$-$y$ centered at the crack tip, the stresses at a distance $r$ ahead of the crack tip, for $\theta = 0$, and the displacements at a distance $r$ behind the crack tip, for $\theta = \pm \pi$, are:

$$(\sigma_{yy} + i\sigma_{xy})_{\theta=0} = \frac{K}{\sqrt{2\pi r}} r^{i\varepsilon}$$

$$(u_y + iu_x)_{\theta=\pi} - (u_y + iu_x)_{\theta=-\pi} = \frac{8}{E_*} \frac{K r^{i\varepsilon+1/2}}{\sqrt{2\pi}(1+2i\varepsilon)\cosh(\pi\varepsilon)}$$

(4)

where $\cosh(\pi\varepsilon) = (1-\beta^2)^{-1/2}$ and $\dfrac{1}{E_*} = \dfrac{1}{2}\left(\dfrac{1}{\overline{E}_1} + \dfrac{1}{\overline{E}_c}\right)$ or $E_* = (1-\alpha)\overline{E}_1$. The absolute value of the complex stress intensity factor $|K| = \left(K_1^2 + K_2^2\right)^{1/2}$ is related to the energy release rate by the unique relation:

$$G = \frac{1-\beta^2}{E_*}|K|^2$$

(5)

For bi-material interfaces characterized by $\varepsilon = \beta = 0$, the crack tip fields are analogous to those of a homogeneous material, $G = G_I + G_{II}$ and $K = K_I + iK_{II} = Ke^{i\psi}$ with the mode mixity phase angle uniquely defined as:

$$\psi = \tan^{-1}(K_{II} / K_I) \quad \text{for } \varepsilon = \beta = 0$$

(6)

---

[1] While deriving Eqs. (2),(3), we observed that the bending stiffness $I_3$ in [2], which is the analogue of our $D_b$, should be modified to $I_3/3$; this change yields the expected limiting result for a homogeneous edge-cracked layer with $\Sigma = \overline{E}_1/\overline{E}_c = 1$ [11].

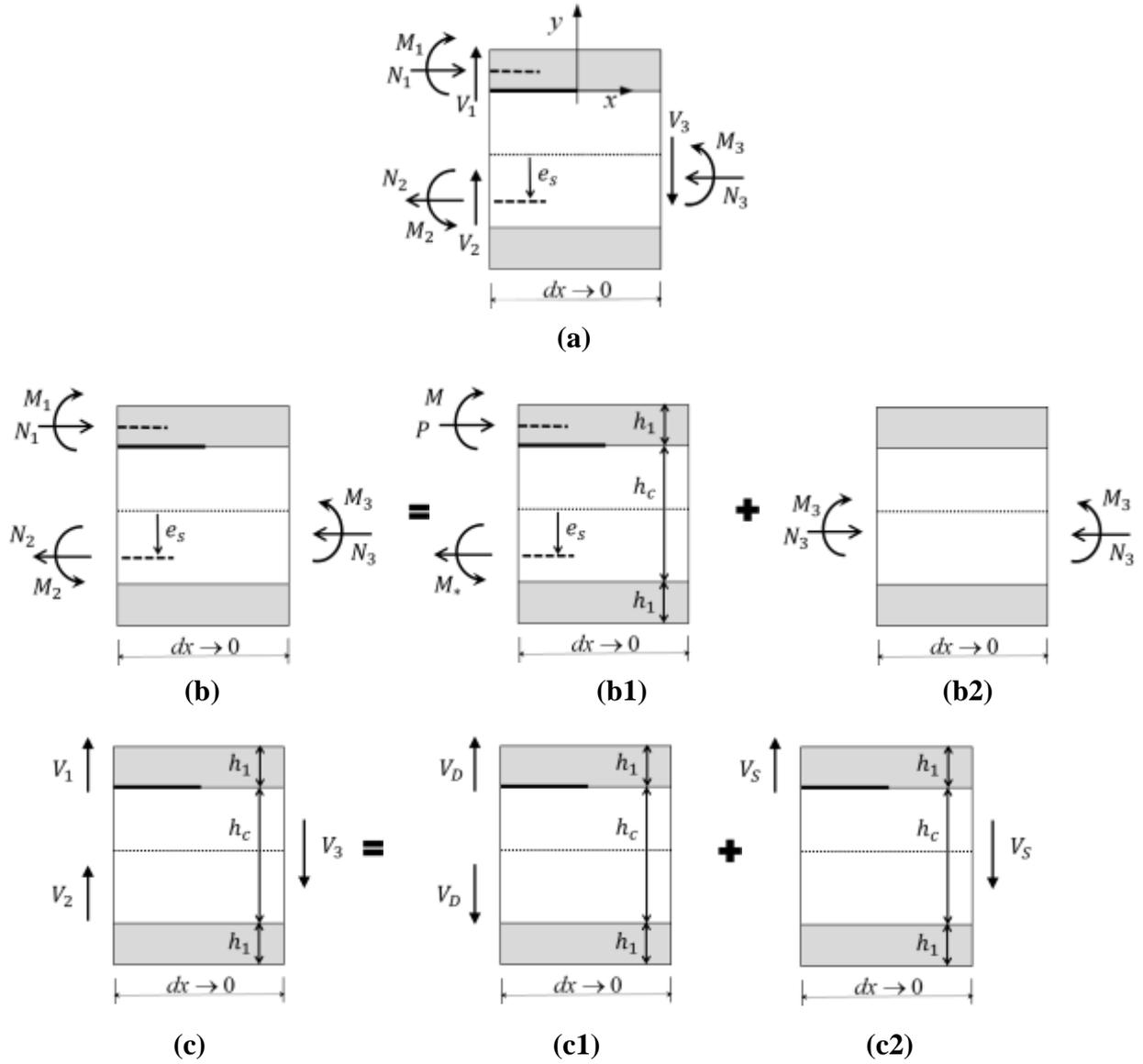

Figure 2. (a) Crack tip loads. (b) Load decomposition bending moments and axial forces: (b1) bending moments, axial forces and compensating bending moment, $M_* = M + P(h_1/2 + h_c/2 + e_S)$; (b2) equilibrating forces acting on the intact crack tip element. (c) Load decomposition shear forces: (c1) double-shear, (c2) single-shear.

In all other cases, the presence of the oscillatory terms in Eqs. (4) implies that the ratio of normal and shear stress components depend on the distance $\hat{r}$ from the crack tip, namely $\left(\sigma_{xy}/\sigma_{yy}\right)_{r=\hat{r}} = \mathrm{Im}\left[K\hat{r}^{i\varepsilon}\right]/\mathrm{Re}\left[K\hat{r}^{i\varepsilon}\right]$. As a consequence the energy release rates for the individual modes do not exist, while G still exist and is given by Eq. (5), and a nominal mode mixity angle, $\psi_{\hat{r}}$, is conventionally introduced and related to the complex stress intensity factor $K\hat{r}^{i\varepsilon} = \left|K\hat{r}^{i\varepsilon}\right|e^{i\psi_{\hat{r}}}$, with $\hat{r}$ a characteristic length of the problem [14]. The absolute value of $K\hat{r}^{i\varepsilon}$ coincides with that of $K$,

$|K\hat{r}^{i\varepsilon}| = |K|$, and its phase angle $\psi_{\hat{r}} = \tan^{-1}\left(\text{Im}\left[K\hat{r}^{i\varepsilon}\right]/\text{Re}\left[K\hat{r}^{i\varepsilon}\right]\right)$ define the ratio of shear and normal stresses at a fixed distance $\hat{r}$ ahead of the crack tip. The characteristic length $\hat{r}$ is assumed in this paper as $\hat{r} = h_1$, as it is usually done for interfacial cracks in beam type geometries [14] [12]. The mode mixity angle which will be derived in this paper for the geometry in Fig. 1 is then:

$$\psi = \psi_{h1} = \tan^{-1}\left(\text{Im}\left[Kh_1^{i\varepsilon}\right]/\text{Re}\left[Kh_1^{i\varepsilon}\right]\right), \text{ for } \varepsilon \neq \beta \neq 0 \tag{7}$$

The definition of the characteristic distance is irrelevant since a simple transformation rule can be derived which relates $\psi = \psi_{h1}$ and $\psi_{\hat{r}}$ at a generic distance $\hat{r}$, namely $\psi_{\hat{r}} = \psi_{h1} + \varepsilon \ln(\hat{r}/h_1)$ [6]. The mode mixity angle calculated using Eq. (7) can then be used to define the mode mixity angle at different distances. The relationship between the mode mixity angle as defined above through LEFM and the mode mixity associated to a cohesive description of the problem in small-scale yielding and the definition of a meaningful characteristic length, $\hat{r}$, for the two treatments to be in agreement when $\beta \neq 0$, is still an open problem [16], [17].

The presence of the oscillatory term in Eq. (4) also implies the existence of a region at the crack tip where the crack faces interpenetrate. The interpenetration is inadmissible and in the actual body would be avoided by the development of a region where the crack faces are in contact [18]. Equation (4) will then accurately describe the displacement field at sufficient distance from the crack tip, only if the region of contact is small compared to any characteristic lengths of the problem (*small-scale contact*). A rough estimate of the size of the contact region is obtained by calculating the distance, $r_c$, at which the relative displacement $\Delta u_y = (u_y)_{\theta=\pi} - (u_y)_{\theta=-\pi}$ vanishes [14]. Using Eq. (4) and imposing $\text{Re}[K\sqrt{2}r_c^{i\varepsilon}/(1+2i\varepsilon)] = 0$, with $Kh_1^{i\varepsilon} = |K|e^{i\psi}$ and $(1+2i\varepsilon) = \sqrt{1+4\varepsilon^2}\,e^{i\tan^{-1}(2\varepsilon)}$, yields $\cos[\psi - \varepsilon \ln(h_1/r_c) - \tan^{-1}(2\varepsilon)] = 0$ so that:

$$r_c = h_1 \exp\left(-\frac{\pi/2 + \psi - \tan^{-1}(2\varepsilon)}{\varepsilon}\right) \tag{8}$$

In [14] Rice suggests the use of $r_c \leq h_1/100$ as a suitable restriction for the validity of the LEFM solution in the presence of interpenetrating/contact regions, a requirement which is generally met in the presence of some nonnegligible tensile components of the loading relative to the crack.

The relative crack displacements in the singularity dominated zone, Eq. (4), are used in this paper within a correlation technique based on the Crack Surface Displacement (CSDE) method (see

Section 6), to define the mode mixity angle and the energy release rate from accurate FE measurements. Equation (4) can be elaborated as follows:

$$|\Delta u| e^{i\phi_r} = \frac{8}{E_*} \frac{|K| e^{i\psi} r^{i\varepsilon+1/2}}{\sqrt{2\pi} h^{i\varepsilon} \sqrt{1+4\varepsilon^2} e^{i\tan^{-1}(2\varepsilon)} \cosh(\pi\varepsilon)} \qquad (9)$$

where $|\Delta u| = \sqrt{\Delta u_y^2 + \Delta u_x^2}$, $\Delta u_i = (u_i)_{\theta=\pi} - (u_i)_{\theta=-\pi}$ and $\phi_r = \tan^{-1}(\Delta u_x / \Delta u_y)_r$. Eq. (9) shows that $e^{i\psi} = e^{i(\phi_r + \varepsilon \ln h_1 - \varepsilon \ln r + \tan^{-1}(2\varepsilon))}$ and yields a relationship between the mixed mode phase angle $\psi$ and relative crack displacement measures at a distance $r$ from the crack tip:

$$\psi = \psi_{h_1} = \tan^{-1}(\Delta u_x / \Delta u_y)_r - \varepsilon \ln(r/h_1) + \tan^{-1}(2\varepsilon) \qquad (10)$$

The energy release rate is defined using Eqs. (5),(9):

$$G = \frac{(\Delta u_y^2 + \Delta u_x^2)(1+4\varepsilon^2)\pi E_*}{32 r} \qquad (11)$$

## 3. Fracture parameters in sandwich beams subjected to axial forces and bending moments

### 3.1 Energy release rate

The energy release rate in a sandwich beam subjected to axial forces and bending moments only, Fig. 1, is defined using the J-integral calculated between two points on the crack surfaces and along the external boundaries. In terms of the energetically orthogonal force and moment resultants, $P$ and $M$, it takes the form [8]:

$$G_{M+P} = \frac{1}{2}\left(\frac{P^2}{A_d} + \frac{M^2}{D_d}\right) + \frac{1}{2}\left(\frac{P_*^2}{A_s} + \frac{M_*^2}{D_s}\right) \qquad (12)$$

where $A_d, A_s, D_d, D_s$ are the axial and bending stiffnesses of the delaminated and substrate arms, given in Appendix A1, Eqs. (36),(37). Substituting Eqs. (2),(3),(36),(37) into Eq. (12), the contributions of the single load systems can be highlighted as in [6],[2]:

$$G_{M+P} = \frac{1}{\overline{E}_1}\left(f_M^2 \frac{M^2}{h_1^3} + f_P^2 \frac{P^2}{h_1} + 2 f_M f_P \cdot \sin(\gamma_M) \frac{PM}{h_1^2}\right) \qquad (13)$$

where

$$f_M(\Sigma, \eta) = \left[\frac{1}{2}\left(12 + \frac{1}{\tilde{D}_S}\right)\right]^{1/2} \qquad (14)$$

$$f_P(\Sigma,\eta) = \left[\frac{1}{2}\left(1 + \frac{\Sigma\eta}{1+\Sigma\eta} + \frac{1}{\tilde{D}_S}\left(\frac{1}{2} + \frac{1}{2\eta} + \tilde{e}_s\right)^2\right)\right]^{1/2} \qquad (15)$$

$$\gamma_M(\Sigma,\eta) = \sin^{-1}\left[\frac{1}{2\tilde{D}_S f_P f_M}\left(\frac{1}{2} + \frac{1}{2\eta} + \tilde{e}_s\right)\right] \qquad (16)$$

The solution for the limiting configuration of a homogeneous edge cracked layer, with $\Sigma=1$, is presented in the Appendix A1, Eqs. (41)-(45).

### 3.2 Mode mixity phase angle

The absolute values of the complex stress intensity factors associated to the elementary load systems, $P$ and $M$, are defined using Eqs. (5) and (13), $|K_P| = \sqrt{(1-\alpha_1)(1-\beta_1^2)^{-1}}\, f_P P h_1^{-1/2}$ and $|K_M| = \sqrt{(1-\alpha_1)(1-\beta_1^2)^{-1}}\, f_M M h_1^{-3/2}$. The associated mode mixity angles are $\psi_M$ and $\psi_P = \omega$. The stress intensity factors are combined in the complex plane to define the stress intensity factor of the combined load system, $P + M$, as $K_{M+P} h_1^{i\varepsilon} = K_M h_1^{i\varepsilon} + K_P h_1^{i\varepsilon}$, and its mode mixity angle:

$$\psi_{M+P} = \tan^{-1}\left[\frac{\operatorname{Im} K h_1^{i\varepsilon}}{\operatorname{Re} K h_1^{i\varepsilon}}\right] = \tan^{-1}\left[\frac{\lambda \sin(\omega) + \sin(\psi_M)}{\lambda \cos(\omega) + \cos(\psi_M)}\right] \qquad (17)$$

where $\lambda = (f_P P h_1)/(f_M M)$ [6]. In addition, using Eq. (5) for each elementary load cases, yields $G = G_P + G_M + 2(G_P G_M)^{1/2}\cos(\psi_P - \psi_M)$, which can be compared with Eq. (13) to derive a relationship between $\psi_M$ and $\psi_P = \omega$, namely $\cos(\omega - \psi_M) = \sin(\gamma_M)$:

$$\psi_M = \omega + \gamma_M - \pi/2. \qquad (18)$$

This indicates that $\psi_{M+P}$ depends on a single unknown phase angle, which is typically chosen as $\omega$: $\omega$ is load independent and can be calculated from a single FE analysis [6][3]. Eq. (17) can then be modified, by substituting $-\cos(\omega+\gamma_M)$ for $\sin(\psi_M)$ and $\sin(\omega+\gamma_M)$ for $\cos(\psi_M)$, as in [6] [3]. The values of $\omega$ for symmetric sandwich beams, have been derived in [2] for a large range of elastic and geometrical properties. The Tables 1-13 present the values of $\omega$ for the cases examined in this paper obtained from [2], through interpolation/extrapolation of the results in [2] or through accurate FE analyses for configurations not examined in [2]. The uncertainties on the $\omega$ in the Tables

1-13 are ±0.2° (more details on the verification of the accuracy of the tabulated results are given in Appendix A.4)

## 4. Fracture parameters in sandwich beams subjected to shear and arbitrary loadings

Prior investigations on the effects of shear on the fracture parameters in homogeneous orthotropic edge-cracked layers [8] and bi-material layers [7] highlight some important differences with respect to the solutions for beams subjected to axial forces and bending moment only. In the absence of shear forces, Fig. 2b and Eqs. (12), (13), the energy release rate is not affected by the near tip deformations so that the energy release rate calculated using 2D elasticity coincides with predictions of classical lamination plate theory (or Euler Bernoulli beam theory, for homogeneous materials), under the assumption of built-in arms at the crack tip. In the presence of shear, on the other hand, shear deformations along the beam arms and near tip deformations strongly affect the energy release rate, and must be accounted for in the derivation of explicit expressions. Shear deformations are important in short beams and can be neglected in long beams; near tip deformations fundamentally affect the shear-force component of the energy release rate also in long beams [8].

The effects of shear have been accounted for in the evaluation of the energy release rate in edge-cracked orthotropic homogeneous beams in [8], where a semi-analytic expression of *G* has been derived which depends on numerical compliance coefficients introduced to describe the near tip deformations through the so called root-rotations; G has then been used to partition the modes of fracture and define stress intensity factors and mode mixity phase angle. The solutions in [8] depend on a total of five numerically-derived constants. A different method was proposed in [7] to analyze isotropic bi-material layers. The method uses dimensional considerations, linearity, the relationship between energy release rate and the absolute value of the complex stress intensity factors and operations in the complex plane. The solution still depend on five numerically-derived constants. The two approaches yield the same results but offer different advantages in the solution of the problem. In the following they will be integrated in the solution of the sandwich problem.

### 4.1 Elementary loads: single and double shear

Following [7] and through dimensional considerations, the energy release rate in a sandwich beam subjected to the elementary load systems of symmetric (double) and asymmetric (single) shear, Fig. 2b, take the forms:

$$G_{VD} = f_{VD}(\alpha,\beta,\eta)^2 \frac{V_D^2}{\bar{E}_1 h_1} \tag{19}$$

$$G_{VS} = f_{VS}(\alpha,\beta,\eta)^2 \frac{V_S^2}{\overline{E}_1 h_1}$$

where the dimensionless functions, $f_{VD}, f_{VS}$, which are analogous to the functions $f_M, f_P$ for bending and normal forces, Eqs. (14) and (15), depend also on the Dundurs' parameter $\beta$, as it was already observed in [7] for a bi-layer. Since transverse shear is always coupled to bending moments, for global equilibrium, the crack tip conditions of single and double shear can be obtained in the absence of $M$ and $P$ only for the special loading configurations shown in Fig. 3a,b. These configurations have been used for the derivation of the functions $f_{VD}, f_{VS}$ through accurate FE analyses using the method presented in Section 6. The functions are presented in Tables 1-11 on varying $\alpha = 0 \div 0.998$, $\beta = 0 \div 0.4$ and $\eta = 0.01 \div 1$. The tables complement results presented in [8] for homogeneous edge-cracked layers and in [7] for bi-material layers. The Tables 1-9 refer to $\eta = 0.025 \div 1$, $\alpha = 0 \div 0.8$ and $\beta = 0 \div 0.4$, and describe three monolithic composite layers. The Tables 10-13 refer to $\eta = 0.01 \div 0.15$, $\alpha = 0.8 \div 0.998$ and $\beta = 0.2 \div 0.4$, which include most sandwich systems with very compliant cores (results in Tables 10-13 are presented only for $\beta \geq 0.2$ to describe actual materials with typical Poisson ratios and are limited to sandwiches with thin face sheets).

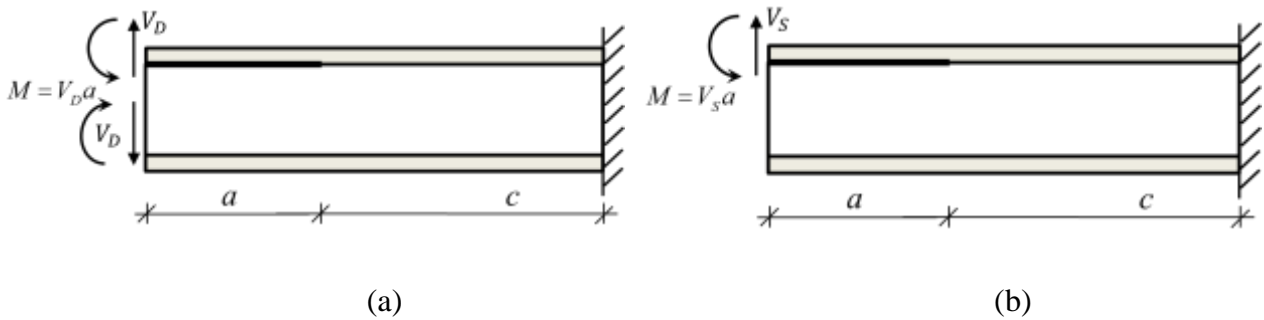

(a)                            (b)

Figure 3. Schematics used for the calculation of energy release rates and mode mixity angles associated to the crack tip conditions of (a) double shear, Fig. 2c1, and (b) single shear, Fig. 2c2.

The complex stress intensity factors associated to the elementary shear loading cases in Fig. 2c, are $K_{VD} h_1^{i\varepsilon} = |K_{VD}| e^{i\psi_{VD}}$ and $K_{VS} h_1^{i\varepsilon} = |K_{VS}| e^{i\psi_{VS}}$, with moduli $|K_{VD}|$ and $|K_{VS}|$ related through Eq. (5) to the energy release rate, Eq. (19), and phase angles, $\psi_{VD} = \tan^{-1}\left(\text{Im}\left[K_{VD} h_1^{i\varepsilon}\right] / \text{Re}\left[K_{VD} h_1^{i\varepsilon}\right]\right)$ and $\psi_{VS} = \tan^{-1}\left(\text{Im}\left[K_{VS} h_1^{i\varepsilon}\right] / \text{Re}\left[K_{VS} h_1^{i\varepsilon}\right]\right)$. The phase angles calculated numerically for a bi-layer and a homogeneous ortotropic edge-cracked layer are presented in [7] and [8]. The Tables 1-13 presents numerical predictions obtained using the method in Section 6 for symmetric sandwich beams. The Tables also include the phase angle $\omega$ for the elementary load case $P$, derived from [2] or through specific FE analyses.

## 4.2. Solutions for arbitrary end forces

The complex stress intensity factors corresponding to the elementary load cases are summed up in the complex plane, Fig. 4, to define the stress intensity factor for general loading cases with $P, M, V_D, V_S \neq 0$. The complex stress intensity factor has modulus $|K|$ and phase angle $\psi = \psi_{h_1} = \tan^{-1}\left(\text{Im}\left[Kh_1^{i\varepsilon}\right]/\text{Re}\left[Kh_1^{i\varepsilon}\right]\right)$. Using the relationship between $|K|$ and G in Eq. (13) and Fig. 4 yields:

$$G = \frac{f_M^2 M^2}{\bar{E}_1 h_1^3} + \frac{f_P^2 P^2}{\bar{E}_1 h_1} + 2\frac{f_M f_P}{\bar{E}_1 h_1^2}\sin\gamma_M MP \qquad (20)$$
$$+ \frac{f_{VD}^2 V_D^2}{\bar{E}_1 h_1} + \frac{f_{VS}^2 V_S^2}{\bar{E}_1 h_1} + 2\frac{f_{VD} f_{VS}}{\bar{E}_1 h_1}\cos(\psi_{VD} - \psi_{VS}) V_D V_S$$
$$+ 2\frac{f_M f_{VD}}{\bar{E}_1 h_1^2}\cos(\psi_M - \psi_{VD}) MV_D + 2\frac{f_M f_{VS}}{\bar{E}_1 h_1^2}\cos(\psi_M - \psi_{VS}) MV_S$$
$$+ 2\frac{f_P f_{VD}}{\bar{E}_1 h_1}\sin\gamma_{VD} PV_D + 2\frac{f_P f_{VS}}{\bar{E}_1 h_1}\sin\gamma_{VS} PV_S.$$

where $\psi_M = \omega + \gamma_M - \pi/2$, $\gamma_{VD} = \psi_{VD} - \omega + \pi/2$, $\gamma_{VS} = \psi_{VS} - \omega + \pi/2$; $\gamma_M, f_P, f_M$ are given in Eqs. (14)-(16), and the numerical coefficients, $f_{VD}, f_{VS}, \psi_{VD}, \psi_{VS}, \omega$, are given in the Tables 1-13.

The mode mixity angle takes the form:

$$\psi = \psi_{h_1} = \tan^{-1}\left(\frac{f_M M \sin\psi_M + f_P P h_1 \sin\omega + f_{VD} V_D h_1 \sin\psi_{VD} + f_{VS} V_S h_1 \sin\psi_{VS}}{f_M M \cos\psi_M + f_P P h_1 \cos\omega + f_{VD} V_D h_1 \cos\psi_{VD} + f_{VS} V_S h_1 \cos\psi_{VS}}\right) \qquad (21)$$

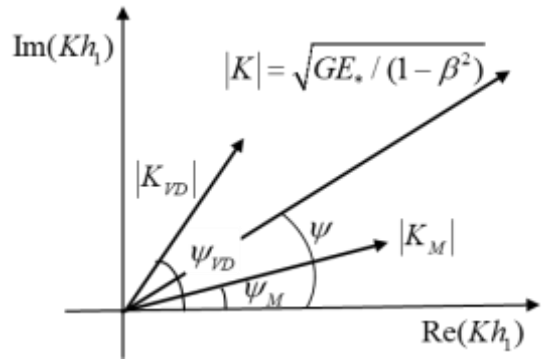

Figure 4. Operations in the complex plane of the stress intensity factors, exemplary application to problem with $M, V_D \neq 0$ and $P, V_S = 0$, e.g. a DCB sandwich specimen.

## 4.3 Physical and mechanical significance of the terms and numerical coefficients in the energy release rate

Equations (20),(21) and the Tables 1-13, can be applied to accurately define the energy rate in sandwich beams with traction free interface cracks subjected to arbitrary loading conditions. However, the physical and mechanical significance of the numerical coefficients and the different terms, in particular the coupling terms which connect different elementary loads in Eq. (20), is not apparent. To overcome this problem, the method formulated in [8] to define the energy release rate of edge-cracked homogeneous orthotropic layers will be extended here to sandwich beams. The method uses an approximate description of the shear contributions, which is based on structural mechanics concepts. Following [8], the primary effect of the near tip deformations on the fracture parameters is accounted for through the introduction of root rotations, which describe the relative rotations of the beam segments 1-3 (delaminated-base) and 2-3 (substrate-base) at the crack tip cross sections:

(22)
$$\Delta\varphi_{3,1}(\alpha,\beta,\eta,P,M,V_S,V_D) = \varphi_3 - \varphi_1$$
$$\Delta\varphi_{3,2}(\alpha,\beta,\eta,P,M,V_S,V_D) = \varphi_3 - \varphi_2$$

with $\varphi_1, \varphi_2, \varphi_3$ the rotations of the cross sections of the debonded, base and substrate arms, Fig. 5a. The rotations are assumed to be positive if counterclockwise. With this assumption the root rotations generated in a Double Cantilever Beam specimen loaded by opening transverse forces F, Fig. 7, will be $\Delta\varphi_{3,1} > 0$ and $\Delta\varphi_{3,2} < 0$, since $\varphi_1 < 0$, $\varphi_2 > 0$ and $\varphi_3 = 0$.

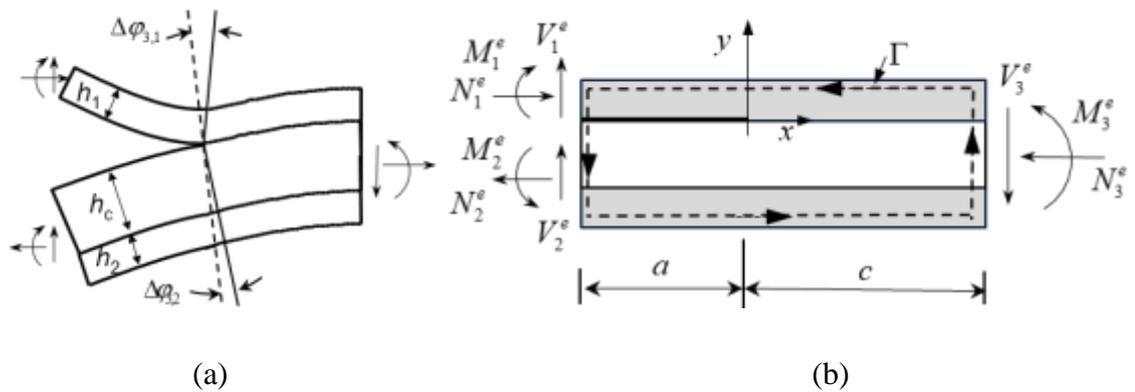

(a) (b)

Figure 5. (a) Root rotations at the crack tip of a sandwich beam. (a) Calculation of the energy release rate in sandwich beam (b)

The root rotations depend on the compliance of the material at and ahead of the crack tip and, if $a$ and $c$ are sufficiently long, can be described as linear combinations of the crack tip force and moment resultants in Fig.2b,c:

$$\Delta\varphi_{3,1} = \frac{1}{\bar{E}_1 h_1}\left(\frac{a_1^M}{h_1}M + a_1^P N + a_1^{VS}V_S + a_1^{VD}V_D\right) \quad (23)$$

$$\Delta\varphi_{3,2} = \frac{1}{\bar{E}_1 h_1}\left(\frac{a_2^M}{h_1}M + a_2^P N + a_2^{VS}V_S + a_2^{VD}V_D\right)$$

where the coefficients $a_i^M, a_i^P, a_i^{VS}, a_i^{VD}$ for i=1,2 define the relative rotations between the delaminated/substrate arms and the bonded arm generated by unit bending moments, axial forces and shear force. The coefficients can be defined through approximate structural models or accurate numerical analyses, as in [8],[21]. In this paper they will be derived by the numerical coefficients which describe the fracture parameters in Eqs. (20) and are given in Tables 1-13.

In beams subjected to bending moments and normal forces only, the near tip deformations and the root rotations, while present, do not affect the energy release rate, $G_{M+P}$ Eq. (12) [12]. Their effect is instead important on the shear components of the energy release rate in beams subjected to arbitrary loading conditions. In order to understand these effects, another expression of the energy release rate is obtained through an application of the J integral along a path which follows the external boundaries of the beam shown in Fig. 5b:

$$G = J = \frac{1}{2}\left[\sum_{i=1}^{2}\left(\frac{(M_i^e)^2}{D_i} + \frac{(V_i^e)^2}{D_{Vi}} + \frac{(N_i^e)}{A_i} - 2V_i\varphi_i^e\right) - \frac{(M_3^e)^2}{D_3} - \frac{(V_3^e)^2}{D_{V3}} - \frac{(N_3^e)^2}{A_3} + 2V_3\varphi_3^e\right] \quad (24)$$

where $i = 1 = d$, $i = 2 = s$, $i = 3 = b$ and $\varphi_i^e$ is the rotation of the end cross section of the arm $i$ [8]. The terms related to the bending moments and axial forces, which depend on the axial and bending stiffnesses of the various arms, given in Eqs. (35)-(38), coincide with the classical solution of 2D elasticity provided $a$ and $c$ are sufficiently long, Eq. (2). The terms related to the shear forces have been obtained using the simplifying kinematic assumption, $\partial v_{yi}/\partial x = \gamma_i + \varphi_i$, which relates transverse displacements, $v_{yi}$, rotations, $\varphi_i$, and global shear strains, $\gamma_i = V_i/D_{Vi}$ (Timoshenko beam theory) in the reference system in Fig. 4b; the global strain is related to the shear force through a shear stiffness, $D_{Vi}$, which must depend on the shear moduli and thicknesses of the different layers and on the thickness-wise shear stress distributions in the arm $i$ (more details on $D_{Vi}$ will be given later in the paper). The terms, $2V_i\varphi_i^e$ for i=1,2 and $2V_3\varphi_3^e$, depend on the rotations of the end cross sections of the beam arms. These rotations can be defined using the virtual work principle and are defined as the sum of the rotations of the crack tip cross sections, $\varphi_1, \varphi_2, \varphi_3$, which describe the near tip deformations, and the rotations due to the applied generalized forces acting on each beam segment end:

$$\varphi_1^e = \varphi_1 - \frac{M_1^e a}{D_1} - \frac{V_1^e a^2}{2D_1}, \quad \varphi_2^e = \varphi_2 + \frac{M_2^e a}{D_2} - \frac{V_2^e a^2}{2D_2} \text{ and } \varphi_3^e = \varphi_3 + \frac{M_3^e a}{D_3} - \frac{V_3^e a^2}{2D_3}. \tag{25}$$

Using Eqs. (22),(25) and expressing the generalized end forces in terms of crack tip force and moment resultants yields:

$$\mathcal{G} = \frac{1}{2}\left[\sum_{i=1}^{2}\left(\frac{(N_i)}{A_i} + \frac{(M_i)^2}{D_i} + \frac{(V_i)^2}{D_{Vi}} + 2V_i\Delta\varphi_{3,i}\right) - \frac{(N_3)^2}{A_3} - \frac{(M_3)^2}{D_3} - \frac{(V_3)^2}{D_{V3}}\right] \tag{26}$$

The equation above highlights clearly the contributions of shear to the energy release rate: the crack tip shear forces act on the shear deformations (the terms with the squared $V_i^2$) and on the root rotations (the terms with the $V_i$). If the shear stiffness of the arms is assumed to be infinite, as in Euler Bernoulli beam theory, than the first terms vanish and the only effect of the shear forces is through their product with the root rotations. This assumptions has been used in [21].

Recalling the superposition in Fig. 2b,c and the relationships (2) and operating some algebraic manipulations to express (26) in terms of elementary loads, $P, M, V_S, V_D$, yields:

$$\mathcal{G} = \frac{f_M^2 M^2}{\overline{E}_1 h_1^3} + \frac{f_P^2 P^2}{\overline{E}_1 h_1} + \frac{MP}{\overline{E}_1 h_1^2} 2 f_M f_P \cdot \sin(\gamma_M) \tag{27}$$
$$+ \frac{V_S^2}{\overline{E}_1 h_1}\left(a_1^{VS} + \frac{1}{2}\left(\frac{1}{\tilde{D}_{Vd}} - \frac{1}{\tilde{D}_{Vb}}\right)\right) + \frac{V_D^2}{\overline{E}_1 h_1}\left(a_1^{VD} - a_2^{VD} + \frac{1}{2}\left(\frac{1}{\tilde{D}_{Vd}} + \frac{1}{\tilde{D}_{Vs}}\right)\right)$$
$$+ \frac{V_S P}{\overline{E}_1 h_1} a_1^P + \frac{V_S M}{\overline{E}_1 h_1^2} a_1^M + \frac{V_D P}{\overline{E}_1 h_1}\left(a_1^P - a_2^P\right) + \frac{V_D M}{\overline{E}_1 h_1^2}\left(a_1^M - a_2^M\right)$$
$$+ \frac{V_D V_S}{\overline{E}_1 h_1}\left(a_1^{VD} + a_1^{VS} - a_2^{VS} + \frac{1}{\tilde{D}_{Vd}}\right)$$

with $\tilde{D}_{Vd} = D_{Vd}/\overline{E}_1 h_1$, and $\tilde{D}_{Vs} = D_{Vs}/\overline{E}_1 h_1$ the dimensionless shear stiffnesses of the beam arms (see Appendix A3).

The two expressions of the energy release rate in Eqs. (27) and (20) can now be compared to define the relationships between the root rotation coefficients, $a_i^M, a_i^P, a_i^{VS}, a_i^{VD}$, for i=1,2, and the dimensionless coefficients of Eq. (20), given in Table 1-13. The terms which multiply by $V_S M$ and $V_S P$ in the two expressions of G,

$$2\cos(\gamma_M - \gamma_{VS}) f_M f_{VS} = a_1^M \tag{28}$$
$$\sin \gamma_{VS} 2 f_P f_{VS} = a_1^P,$$

define the root rotations of the delaminated arm, $\Delta\varphi_{3,1} = a_1^M$ and $\Delta\varphi_{3,1} = a_1^P$, due to unit bending moments $M/(\bar{E}_1 h_1^2) = 1$ and unit normal forces $P/(\bar{E}_1 h_1) = 1$, respectively; these rotations affects the energy release rate in the presence of single-shear. The terms which multiply by $V_D M$ and $V_D P$ define the differences between the root rotations of the delaminated and substrate arms, $a_1^M - a_2^M$ and $a_1^P - a_2^P$, generated respectively by unit bending moments and unit normal forces,:

$$2\cos(\gamma_M - \gamma_{VD}) f_M f_{VD} = a_1^M - a_2^M \tag{29}$$
$$\sin \gamma_{VD} 2 f_N f_{VD} = a_1^P - a_2^P$$

Finally, the terms which multiply by $V_D^2$ and $V_S^2$ define the two primary effects of shear: root rotations $a_1^{VD} - a_2^{VD}$, for unit double-shear, and $a_1^{VS}$ for unit single- shear, and global shear strains, $\gamma_d = (V_D + V_S)/D_{Vd}$, $\gamma_S = V_D/D_{Vs}$ and $\gamma_b = V_S/D_{Vb}$, which enter in the terms depending on the shear stiffnesses:

$$f_{VD}^2 = a_1^{VD} - a_2^{VD} + \frac{1}{2}\left(\frac{1}{\tilde{D}_{Vd}} + \frac{1}{\tilde{D}_{Vs}}\right) \tag{30}$$
$$f_{VS}^2 = a_1^{VS} + \frac{1}{2}\left(\frac{1}{\tilde{D}_{Vd}} - \frac{1}{\tilde{D}_{Vb}}\right)$$

The last relationship is between the terms which multiply by $V_D V_S$:

$$2 f_{VD} f_{VS} \cos(\psi_{VD} - \psi_{VS}) = a_1^{VD} + a_1^{VS} - a_2^{VS} + \frac{1}{\tilde{D}_{Vd}} \tag{31}$$

The equations (28) and (31) can be used to define the root rotation coefficients which appear in Eq. (23) from the dimensionless coefficients in the Tables 1-13 or viceversa.

The two formulas in Eq. (30) show that the functions $f_{VD}^2$ and $f_{VS}^2$ which appear in the expression (20) of the energy release rate, account for different effects of shear: the root rotations due to unit double-shear, $(a_1^{VD} - a_2^{VD})$, and unit single-shear, $a_1^{VS}$, and the shear strains which arise along the beam arms and depend on their normalized shear stiffnesses, $\tilde{D}_{Vs}, \tilde{D}_{Vd}, \tilde{D}_{Vb}$. If the shear stiffnesses of the arms are assumed to be infinite, as in Euler Bernoulli beam theory (or classical plate theory) then all terms with $D_{Vi}$ would vanish in the Eqs. (24)-(31) and the shear strain contributions will be included in the root rotation coefficients obtained through the equivalence (30); this method has been recently applied in [21]. Otherwise, following [8], the shear stiffnesses of the arms may be accounted for so that the effects of the root rotations and those of the shear strains are kept separated in the expression of G. The shear stiffnesses can then be defined through accurate numerical analyses or using the Jourawsky approximation, through the introduction of shear correction factors, $\kappa_{Vi}$, so that $D_{Vi} = \kappa_{Vi} S_i$, with $S_i$ the shear areas. This second approach is presented in the Appendix A3, Eq. (47),(49), and will be used in the exemplary application in Section 5.

**4.4 Minimum lengths of the sandwich arms**

Equations (20),(21) and (27) are applicable to sandwich beams and fracture specimens under arbitrary loading conditions provided the lengths of the traction free crack and intact ligament, *a* and *c*, are sufficiently long to ensure the accuracy of the analytical expressions and the tabulated coefficients, Table 1-13, which have been derived for beams where: (i) the fracture parameters are unaffected by the actual distribution of the applied tractions at the beam ends, and (ii) the stresses from the crack tip do not interact with the beam ends.

In relation to (i), the loads at the beam ends in the FE models in Fig. 4.a,b have been applied using constrained conditions which enforce the nodes to remain on a straight line (see Section 6) and the length of the crack chosen long enough to ensure a constant distribution of shear stresses along the beam arms up to the crack tip. Fig. 6a show a zoom around the crack tip of the dimensionless shear stresses, $\sigma_{xy} h_1 / V_D$, which meet this requirement in the region surrounding the crack tip of a sandwich with three equal thickness layers and a very large mismatch of the Young's moduli, $\eta = 1$, $\Sigma = \bar{E}_1 / \bar{E}_c = 1000$ (or $\alpha = 0.998$) and $\beta = 0.4$, subjected to double shear, $V_D$.

In relation to (ii), it was observed that, as for axial loadings [2], the stresses from the crack tip affect a region which increases in size on increasing the mismatch of the elastic constants of the layers and elongates mainly ahead of the tip; this is evident in the Figs. 6a,b showing shear and transverse stresses in the specimen described above. It was also observed that an important effect of the shear forces acting at the crack tip cross sections in beams with a large mismatch of the Young's moduli, e.g. $\Sigma = \bar{E}_1/\bar{E}_c = 1000$, is to generate bending stresses in the face sheets ahead of the crack tip which follow the typical pattern observed in beams on elastic foundations. This is shown in Figs. 6c,d. These stresses may still be important also at a large distance from the crack tip and if they interact with the boundaries may affect predictions of energy release rate and mode mixity.

In sandwich composites with soft cores and a large mismatch of the elastic constants, $\alpha \geq 0.8$ and the minimum lengths are substantially larger than those which necessitate for conventional composites with $\alpha \leq 0.8$, for which $a_{\min} \approx \max\{h_1, (h_1+h_c)\}\lambda^{-1/4}$ and $c_{\min} \approx (2h_1+h_c)\lambda^{-1/4}$, with $\lambda = E_y/E_x$ in homogeneous orthotropic media and $\lambda = 1$ in isotropic layered media ([7][8][6][2]).

The minimum lengths have been defined in [2] for sandwich beams with $\alpha \geq 0.8$ subjected to axial forces and bending moments as the lengths which ensure relative errors on the energy release rate below 5% for the most severe combination of loads. Under such conditions and for the most critical case of a beam with three equal thickness layers, $\eta = 1$, and $\beta = 0$, the minimum length of the ligament, $c_{min}$, was found be on the order $c_{\min} \approx 3, 10, 30 h_1$ (or $c_{\min} \approx 1, 3, 10(2h_1+h_c)$) for face sheet/core stiffness ratios, $\Sigma = \bar{E}_1/\bar{E}_c = 10, 100, 1000$ (or $\alpha = 0.81, 0.98, 0.998$), while the minimum length of the crack to be on the order $a_{\min} \approx 4h_1$ or $a_{\min} \approx 2(h_1+h_c)$. For sandwiches with $\eta < 1$ the minimum lengths of the ligament progressively reduce and for $\eta = 0.125$ they are $c_{\min} \approx 8, 10, 17 h_1$ (or $c_{\min} \approx 0.8, 1, 1.7(2h_1+h_c)$), similar to those defined for conventional composites with $\alpha \leq 0.8$.

To define the minimum lengths in sandwich beams subjected to shear, the geometry in Fig. 3a has been examined and the effects of progressive reductions of $a$ and $c$ on energy release rate and mode mixity have been investigated. As for the axial/bending case, the most critical case was found to correspond to sandwiches with equal thickness layers, $\eta = 1$; however, for shear, the effect of $\beta$ appears to be opposite and the most critical cases correspond to the largest $\beta = 0.4$. For $\Sigma = \bar{E}_1/\bar{E}_c = 1000$ (or $\alpha = 0.998$) and $\beta = 0.4$ the percent error on both energy release rate and mode mixity angle was found to be lower than 3% already for $c_{\min} \approx 9h_1$ (or $c_{\min} \approx 3(2h_1+h_c)$. The results then ensure that the minimum lengths defined in [2] for axial and bending loads are more than adequate also for problems in the presence of shear forces.

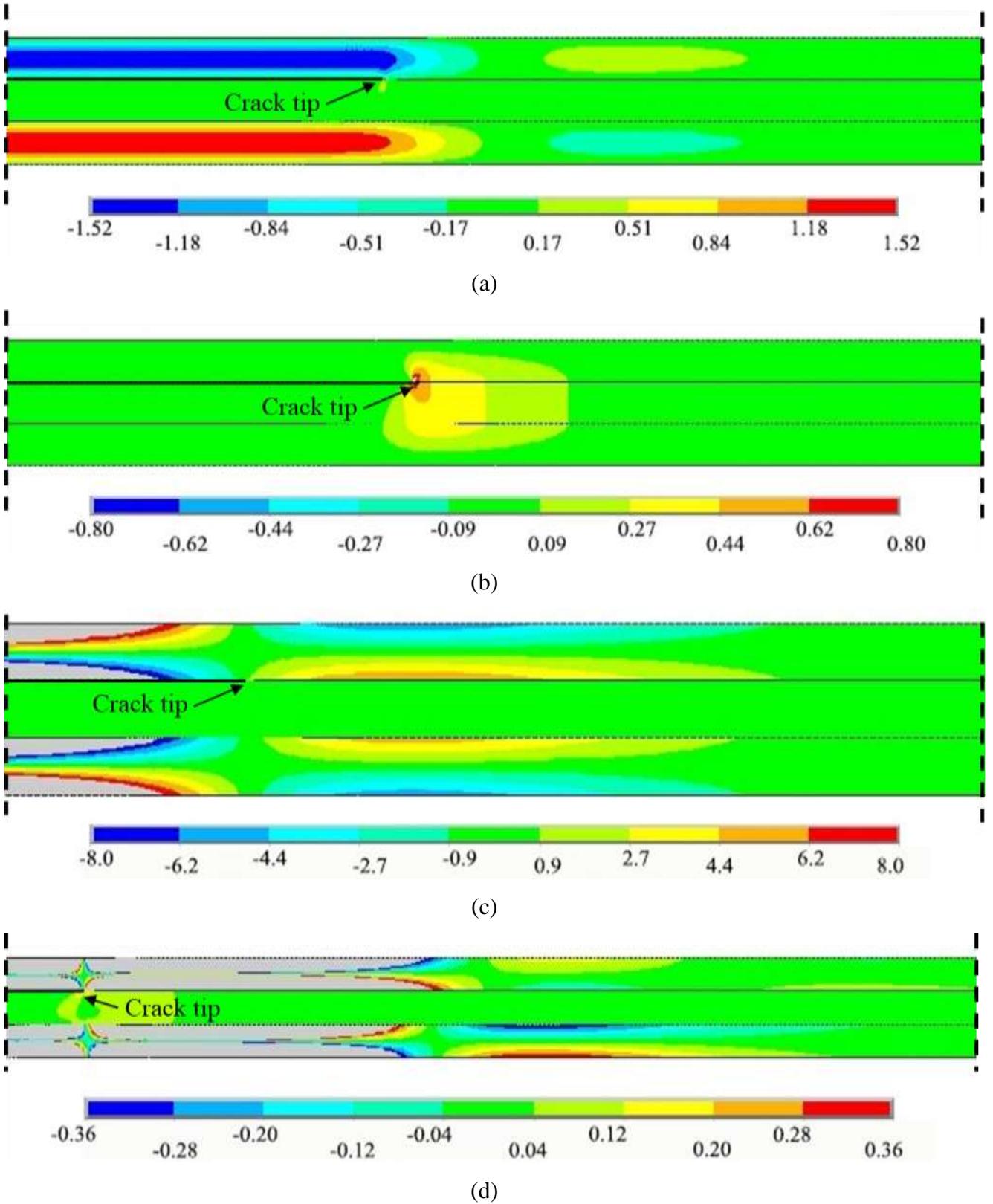

Figure 6. Dimensionless stresses $\sigma_{ij} h_1 / V_D$ in the region surrounding the crack tip of a sandwich specimen with three equal thickness layers and a large mismatch of the elastic constants, subjected to double shear, $V_D$ ($\eta = 1$, $\Sigma = \bar{E}_1 / \bar{E}_c = 1000$ (or $\alpha = 0.998$) and $\beta = 0.4$, $a/h_1 = 25$, $c/h_1 = 25$).

(a) Transverse shear stresses, (b) Transverse normal stresses, (c,d) Bending stresses with typical distribution observed in beams on elastic foundations. Regions in light grey identify positive (negative) values larger (smaller) than those listed in the captions.

## 5. Application to a Double Cantilever Beam sandwich specimen

In a DCB sandwich specimen subjected to transverse forces F at the beam ends, the elementary load systems, Fig. 7, are $M = Fa$, $V_D = F$ while $P, V_S = 0$. The energy release rate and mode mixity angle, Eqs. (20),(21) simplify as:

$$G_{DCB} = \frac{F^2}{\bar{E}_1 h_1}\left[\left(\frac{a}{h_1}\right)^2 f_M^2 + 2\left(\frac{a}{h_1}\right) f_{VD} f_M \cos(\psi_M - \psi_{VD}) + f_{VD}^2\right] \tag{32}$$

$$\psi_{DCB} = \tan^{-1}\left(\frac{a/h_1\, f_M \sin(\psi_M) + f_{VD}\sin(\psi_{VD})}{a/h_1\, f_M \cos(\psi_M) + f_{VD}\cos(\psi_{VD})}\right) \tag{33}$$

where $\psi_M = \omega + \gamma_M - \pi/2$ and $f_M$ and $\gamma_M$ are given by the analytical Eqs. (14) and (15); $f_{VD}$, $\psi_{VD}$ and $\omega$ are given in the numerical Tables 1-11. Using the method presented in Section 4.2 and Eq. (27), G may also be defined by introducing root rotations and shear strains to highlight the physical significance of the different contributions:

$$G_{DCB} = \frac{F^2}{\bar{E}_1 h_1}\left[\left(\frac{a}{h_1}\right)^2 f_M^2 + \left(\frac{a}{h_1}\right)(a_1^M - a_2^M) + \left(a_1^{VD} - a_2^{VD} + \frac{1}{2}\left(\frac{1}{\tilde{D}_{Vd}} + \frac{1}{\tilde{D}_{Vs}}\right)\right)\right] \tag{34}$$

Equations (32),(34) have terms which depend on the crack length at different powers. The dominant term, with the crack length squared, is the contribution of the crack tip bending moments. The second term, which depend linearly on the crack length describes the effects of the shear forces on the root rotations generated by the crack tip bending moments, which are given by $(a_1^M - a_2^M)Fa/(E_1 h_1^2)$. The third term, which is independent of the crack length, and therefore become less important on increasing $a/h_1$, describes the effects of the root rotations due to the shear forces, $(a_1^{VD} - a_2^{VD})F/(E_1 h_1)$, and those of the shear strains.

An exemplary application of the results is presented in Figs. 7b,c,d. Assuming $\eta = h_1/h_c = 0.25$, $\Sigma = \bar{E}_1/\bar{E}_c = 9$ or $\alpha = 0.8$ and $\beta = 0$ yields $f_M = 2.49$ and $\gamma_M = 0.159$; the numerical coefficients are obtained from the Table 4: $\omega = 63.5°$, $\psi_M = -17.5°$, $f_{VD} = 3.294$ and

$\psi_{VD} = -8.3°$. The dimensionless energy release rate and mode mixity angle are shown in the diagrams in Fig. 7 on varying the normalized length of the crack. The results are compared with solutions which neglect root rotations and shear deformations. The different effects of shear are highlighted in Fig. 5c where the solid curve define the absolute relative percent error on the energy release rate if the effects of shear are fully neglected, but for the bending moment generated at the crack tip, and G is assumed as $G_M = \dfrac{F^2}{\bar{E}_1 h_1}\left(\dfrac{a}{h_1}\right)^2 f_M^2$. The error decays very slowly and is 40% for $a/h_1 = 5$ and around 10% for $a/h_1 = 30$. The error comes from two contributions. The dominant contribution is that due to neglecting the root rotations produced by the crack tip bending moments (dashed curve). The second contribution is the error due to neglecting shear strains and root rotations generated by shear, which is given by the difference between the two curves (double arrow). The two contributions due to the latter error have been derived in the Appendix A3. It turns out that in a beam with $v_1 = v_c = 0.5$, 44% of the error is due to neglecting the shear root rotations and 56% to neglecting the shear strains. These figures modify, but not substantially, on varying the Poisson ratios while keeping $\beta$ constant and for $v_1 = 0.2$ and $v_c = 0.48$, they become 54% and 46%.

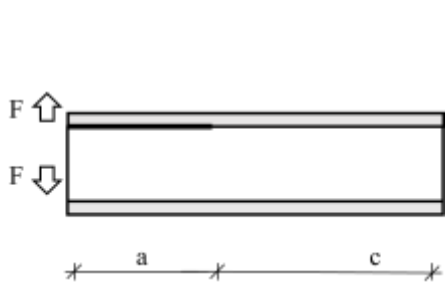

(a)

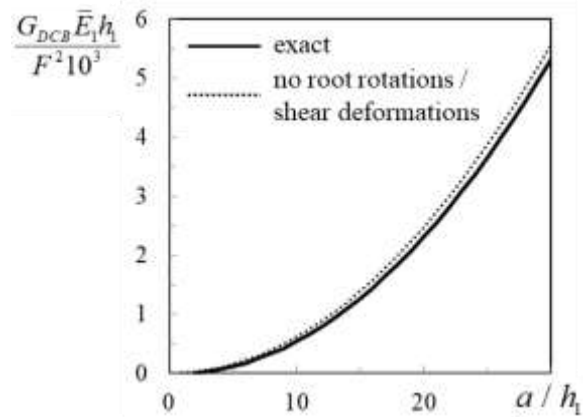

(b)

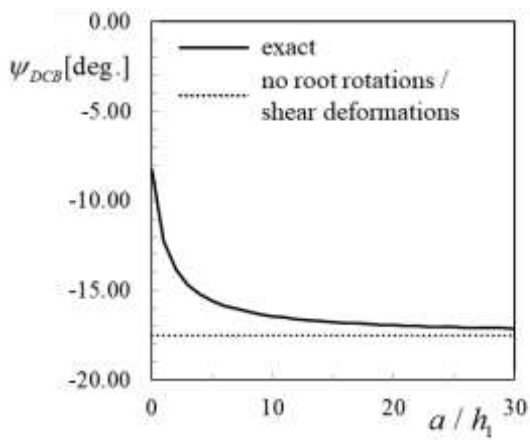

(c)

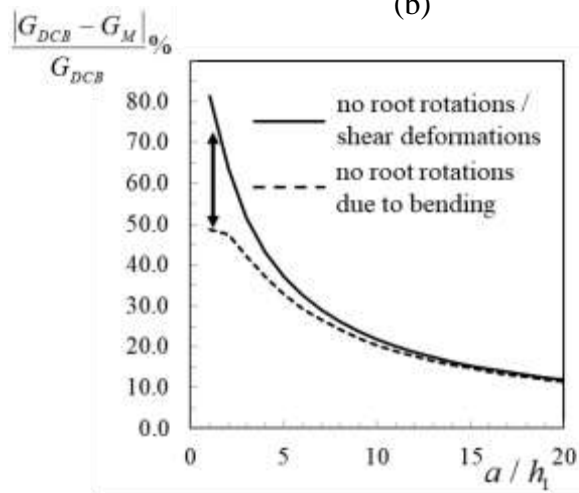

(d)

Figure 7. (a) DCB sandwich specimen with $\eta = h_1/h_c = 0.25$, $\Sigma = \bar{E}_1/\bar{E}_c = 9$ or $\alpha = 0.8$ and $\beta = 0$. (b) Energy release rate: exact solution (solid line), neglecting root rotations and shear deformations (dotted line). (c) Mode mixity angle: exact solution (solid line), neglecting root rotations and shear deformations (dotted line). (d) Absolute percent error on the calculation of the energy release rate: neglecting root rotations and shear deformations (solid line), neglecting root rotations produced by the crack tip bending moments (dashed line), neglecting shear deformations and root rotations due to shear (arrow).

The results of the Eqs. (32),(33) have also been compared with accurate finite element predictions of the experimental DCB test obtained using ANSYS and the CSDE method. For $a/h_1 = 12.5$ or $a/h_c = 3.125$ and $c/h_1 = 12.5$ the relative percent errors between Eqs. (32), (33) and the numerical predictions are 0.4% and 0.18% on energy release rate and mode mixity angle respectively. If the shear contributions are neglected the relative errors go up to 18% and 5.32%. For longer cracks, $a/h_1 = 25$, the effects of shear are still very important and relative differences between the solutions which neglect the effects of shear and the numerical predictions are 9.7% and 2.72%.

## 6. Numerical model for the derivation of the numerical coefficients

The derivation of the dimensionless functions, $f_{VD}, f_{VS}$, which define the energy release rate contributions associated to the elementary loadings of double- and single-shear, and the associated mode mixity phase angles, $\psi_{VD}$ and $\psi_{VS}$, require accurate numerical fracture mechanics calculations, which have been conducted in this paper using the CSDE method formulated in [27] and implemented in the finite element code ANSYS Mechanical.

The CSDE method uses measures of the relative crack displacements along the crack surfaces taken in a process region within the singularity dominated zone and Eqs. (10),(11) to define local values of energy release rate and mode mixity phase angle which are then linearly extrapolated to $r \to 0$. The relative crack displacements are calculated using the finite element method. The typical mesh for the models in Figs. 3a,b have been defined through a mesh convergence study and is subdivided into two domains. The external domain uses 8-noded isoparametric elements (PLANE183) with size not larger than $h_i/8$, for i = 1,2,3 (the three arms). The internal domain is composed of a succession of approximately 100 rings with elements of decreasing size towards the crack tip; the 8 innermost rings surrounding the crack tip are composed of 256, 4-noded, isoparametric elements (PLANE 182) of size $\sim h_1/4000$; the 75 surrounding rings are composed of 8-noded isoparametric elements of size $\sim h_1/1500$. The process region used to calculate the fracture parameters falls into these 83 rings. The mesh has been constructed to keep the maximum aspect ratio of the elements below 10 in order to have a smooth transition between the external and the internal domains and avoid stress discontinuities.

The lengths of the crack and the ligament ahead of the crack tip have been defined in order to ensure that the boundary conditions at the edges do not affect the stress distribution in the crack tip region and that the stresses generated at the crack tip do not interact with the beam ends. Specific values of the limit lengths are given in Section 4.4. The loads have been applied at the free ends using Multi-Point Constraints (MPC elements) which enforce the nodes of the end cross sections to remain on a straight line.

The accuracy of the numerical model used to derive the tabulated dimensionless coefficients has been verified by comparing results for energy release rate and mode mixity angles with analytic elasticity solutions and numerical results presented in the literature and obtained using different derivation methods ([11][7][8],[6][7][12],[2][3]). A detailed presentation of the verification is in Appendix A4.

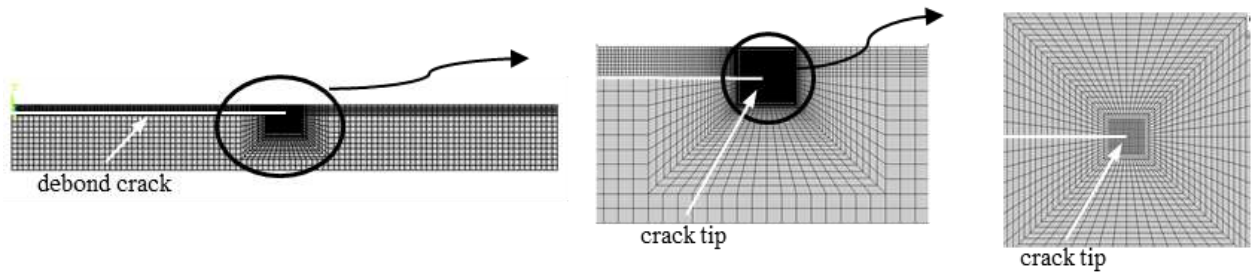

Figure 8. Typical finite element mesh used in ANSYS with the CSDE method to calculate the fracture parameters, showing the external and internal domains.

## 7. Conclusions

Semi-analytic expressions have been derived, using two-dimensional elasticity, dimensional analysis and interface fracture mechanics, for energy release rate and mode mixity phase angle of symmetric sandwich beams with isotropic layers and face-core debonds subjected to arbitrary end forces (Fig. 1). The expressions are presented in Eqs. (20),(21) as functions of the crack tip force and moment resultants, material and geometrical parameters and five numerical coefficients. The coefficients $f_{VD}$ and $f_{VS}$ directly define the energy release rate in beams subjected to the elementary crack tip loads of double- and single-shear, through Eq. (19) (Fig. 2c); the coefficients $\psi_P = \omega, \psi_{VD}, \psi_{VS}$ are the mode mixity angles corresponding to the elementary loads of pure normal forces, plus the required compensating moment (Fig. 2b), double- and single-shear. In the absence of shear forces the expressions coincide with those derived in [2] for symmetric sandwich beams subjected to normal forces and bending moments and a single numerical coefficient, $\psi_P = \omega$, is required.

The coefficients have been derived through accurate finite element analyses and are presented in tabular form in the Tables 1-9 for symmetric sandwiches made of conventional composites with Dundurs' parameters $\alpha = 0 \div 0.8$ (or $\Sigma = \bar{E}_1/\bar{E}_c = 1 \div 9$) and $\beta = 0 \div 0.4$, and relative thicknesses of the layers varying in the range $\eta = h_1/h_c = 0.025 \div 1$. Additional results for symmetric sandwiches with a larger mismatch of the elastic properties, $\alpha = 0.8 \div 0.998$ (or $\Sigma = \bar{E}_1/\bar{E}_c = 9 \div 1000$) and $\beta = 0.2 \div 0.4$, which could describe systems with polymeric foam cores, with $\eta = h_1/h_c = 0.01 \div 0.15$ are presented in Tables 10-13. Linear interpolation between the tabular data can be used to define coefficients for different geometries or material combinations.

The physical and mechanical significance of the numerical coefficients in the expressions of energy release rate and mode mixity is explained using structural mechanics concepts and introducing crack tip root rotations to describes the main effects of the shear and near tip deformations on the fracture parameters in the presence of shear forces. The coefficients $f_{VD}, f_{VS}$ are then directly related to the root rotations produced at the crack tip by unitary double- and single-shear loads, through Eqs. (30), and the mode mixity angles, $\psi_{VD}, \psi_{VS}$, are found to depend on the root rotations produced at the crack tip by all elementary loads, Eqs. (28)(29).

The expressions (20),(21) can be applied to define the fracture parameters of laboratory specimens used for the characterization of the fracture properties of sandwich materials, as shown for the exemplary case of the DCB specimen in Section 5. They can also be applied to complement finite element calculations used to define the crack tip force and moment resultants in beams or wide plates subjected to more complex loading and boundary conditions; the FE analyses will then be used to define the crack tip force and moment resultants, which will then be inserted into Eqs. (20),(21), so avoiding the complex case specific FE calculation of the fracture parameters.

In sandwich specimens with a large mismatch of the elastic properties, $\Sigma = \bar{E}_1/\bar{E}_c = 10 \div 1000$, the solutions accurately define the fracture parameters in actual specimens if the lengths of the crack and the ligament ahead of the crack tip, $a$ and $c$, are sufficiently long. The presence of the crack tip shear forces does not change what observed in [2] for beams subjected to normal forces and bending moments only. For sandwiches with thin face sheets, $\eta = h_1/h_c \leq 0.15$, the minimum lengths are found to be similar to those of conventional composites with $\Sigma = \bar{E}_1/\bar{E}_c = 1 \div 10$ and are on the order ot the heights, namely $a_{min} \approx (h_1 + h_c)$ and $c_{min} \approx 0.8, 1, 1.7(2h_1 + h_c)$ for $\bar{E}_1/\bar{E}_c = 10, 100, 1000$, respectively. The minimum lengths progressively increase on increasing $\eta = h_1/h_c$ to approach $a_{min} \approx 2(h_1 + h_c)$ and $c_{min} \approx 1, 3, 10(2h_1 + h_c)$, for $\bar{E}_1/\bar{E}_c = 10, 100, 1000$, when $\eta = 1$.

**Acknowledgements:** The support of the U.S. Navy, Office of Naval Research, ONR, grants no: N00014-17-1-2914 (Massabò) and N00014-16-1-2977 (Berggreen) monitored by Dr. Y.D.S. Rajapakse is gratefully acknowledged. The first author also acknowledges financial support of the (MURST) Italian Department for University and Scientific and Technological Research in the framework of the research MIUR Prin15 project 2015LYYXA8.

Table 1: Dimensionless functions $f_{VD}(\alpha,\beta,\eta)$, $f_{VS}(\alpha,\beta,\eta)$ and phase angles $\psi_{VD}$, $\omega$ and $\psi_{VS}$, for interface crack in symmetric sandwich with $\eta = h_1/h_c = 0.025$ and $\alpha = (\bar{E}_1 - \bar{E}_c)/(\bar{E}_1 + \bar{E}_c) \leq 0.8$ (additional results for $\alpha \geq 0.8$ in Tables 11-14)

| $\beta$ \ $\alpha$ | \multicolumn{2}{c}{0} | \multicolumn{2}{c}{0.2} | \multicolumn{2}{c}{0.4} | \multicolumn{2}{c}{0.6} | \multicolumn{2}{c}{0.8} |
|---|---|---|---|---|---|---|---|---|---|---|
| | $f_{VD}$ | $f_{VS}$ | $f_{VD}$ | $f_{VS}$ | $f_{VD}$ | $f_{VS}$ | $f_{VD}$ | $f_{VS}$ | $f_{VD}$ | $f_{VS}$ |
| 0   | 1.937 | 1.933 | 2.076 | 2.069 | 2.276 | 2.265 | 2.601 | 2.582 | 2.283 | 3.250 |
| 0.1 | 1.967 | 1.962 | 2.106 | 2.098 | 2.306 | 2.294 | 2.630 | 2.610 | 3.311 | 3.273 |
| 0.2 | 1.992 | 1.989 | 2.132 | 2.123 | 2.331 | 2.318 | 2.655 | 2.632 | 3.333 | 3.290 |
| 0.3 |       |       |       |       | 2.353 | 2.341 | 2.675 | 2.651 | 3.350 | 3.303 |
| 0.4 |       |       |       |       |       |       |       |       | 3.362 | 3.311 |

| $\beta$ \ $\alpha$ | \multicolumn{3}{c}{0} | \multicolumn{3}{c}{0.2} | \multicolumn{3}{c}{0.4} | \multicolumn{3}{c}{0.6} | \multicolumn{3}{c}{0.8} |
|---|---|---|---|---|---|---|---|---|---|---|---|---|---|---|---|
| | $\psi_{VD}$ | $\omega$ | $\psi_{VS}$ | $\psi_{VD}$ | $\omega$ | $\psi_{VS}$ | $\psi_{VD}$ | $\omega$ | $\psi_{VS}$ | $\psi_{VD}$ | $\omega$ | $\psi_{VS}$ | $\psi_{VD}$ | $\omega$ | $\psi_{VS}$ |
| 0   | 0.6°  | 52.2° | 0.4°  | –0.7° | 54.2° | –0.9° | –2.0° | 56.5° | –2.3° | –3.3° | 60.4° | –4.0° | –4.7° | 65.4° | –6.0° |
| 0.1 | –1.9° | 47.8° | –2.0° | –2.9° | 49.9° | –3.1° | –4.0° | 52.3° | –4.3° | –5.0° | 56.3° | –5.6° | –5.8° | 61.4° | –7.1° |
| 0.2 | –4.5° | 43.2° | –4.6° | –5.3° | 45.5° | –5.4° | –6.1° | 48.2° | –6.4° | –6.8° | 52.5° | –7.4° | –7.0° | 57.7° | –8.3° |
| 0.3 |       |       |       |       |       |       | –8.3° | 43.5° | –8.6° | –8.6° | 48.1° | –9.2° | –8.3° | 53.8° | –9.5° |
| 0.4 |       |       |       |       |       |       |       |       |       |       |       |       | –9.8° | 49.1° | –11.0° |

*Interpolation/extrapolation on $\alpha, \beta$ and extrapolation on $\eta$ of the $\omega$ values in [2].

Uncertainties on the tabulated values: $\pm 0.2°$ on $\psi_{VD}$, $\psi_{VS}$; $\pm 0.007$ on $f_{VD}$, $f_{VS}$.

Table 2: Dimensionless functions $f_{VD}(\alpha,\beta,\eta)$, $f_{VS}(\alpha,\beta,\eta)$ and phase angles $\psi_{VD}$, $\omega$ and $\psi_{VS}$, for interface crack in symmetric sandwich with $\eta = h_1/h_c = 0.05$ and $\alpha = (\bar{E}_1 - \bar{E}_c)/(\bar{E}_1 + \bar{E}_c) \leq 0.8$ (additional results for $\alpha \geq 0.8$ in Tables 11-14)

| $\beta$ \ $\alpha$ | \multicolumn{2}{c}{0} | \multicolumn{2}{c}{0.2} | \multicolumn{2}{c}{0.4} | \multicolumn{2}{c}{0.6} | \multicolumn{2}{c}{0.8} |
|---|---|---|---|---|---|---|---|---|---|---|
| | $f_{VD}$ | $f_{VS}$ | $f_{VD}$ | $f_{VS}$ | $f_{VD}$ | $f_{VS}$ | $f_{VD}$ | $f_{VS}$ | $f_{VD}$ | $f_{VS}$ |
| 0   | 1.931 | 1.916 | 2.069 | 2.048 | 2.268 | 2.238 | 2.593 | 2.549 | 3.278 | 3.213 |
| 0.1 | 1.960 | 1.944 | 2.098 | 2.075 | 2.297 | 2.264 | 2.621 | 2.573 | 3.304 | 3.229 |
| 0.2 | 1.986 | 1.969 | 2.123 | 2.099 | 2.331 | 2.287 | 2.645 | 2.592 | 3.325 | 3.240 |
| 0.3 |       |       |       |       | 2.343 | 2.305 | 2.665 | 2.607 | 3.341 | 3.246 |
| 0.4 |       |       |       |       |       |       |       |       | 3.353 | 3.248 |

| $\beta$ \ $\alpha$ | \multicolumn{3}{c}{0} | \multicolumn{3}{c}{0.2} | \multicolumn{3}{c}{0.4} | \multicolumn{3}{c}{0.6} | \multicolumn{3}{c}{0.8} |
|---|---|---|---|---|---|---|---|---|---|---|---|---|---|---|---|
| | $\psi_{VD}$ | $\omega$ | $\psi_{VS}$ | $\psi_{VD}$ | $\omega$ | $\psi_{VS}$ | $\psi_{VD}$ | $\omega$ | $\psi_{VS}$ | $\psi_{VD}$ | $\omega$ | $\psi_{VS}$ | $\psi_{VD}$ | $\omega$ | $\psi_{VS}$ |
| 0   | 0.3°  | 52.2°  | –0.1° | –1.1° | 54.2°* | –1.6° | –2.4° | 56.5°* | –3.3° | –3.9° | 60.4°* | –5.3° | –5.3° | 65.3°* | –7.8° |
| 0.1 | –2.2° | 47.8°  | –2.5° | –3.3° | 49.9°* | –3.8° | –4.4° | 52.3°* | –5.3° | –5.5° | 56.3°* | –6.9° | –6.4° | 61.3°* | –8.9° |
| 0.2 | –4.8° | 43.2°  | –5.1° | –5.6° | 45.5°* | –6.2° | –6.5° | 48.1°* | –7.4° | –7.3° | 52.4°* | –8.7° | –7.6° | 57.6°* | –10.1° |
| 0.3 |       |        |       |       |        |       | –8.7° | 43.5°* | –9.5° | –9.1° | 48.0°* | –10.5° | –8.9° | 53.7°* | –11.3° |
| 0.4 |       |        |       |       |        |       |       |        |       |       |        |       | –10.4° | 49.0°* | –12.7° |

*Interpolation/extrapolation on $\alpha, \beta$ and extrapolation on $\eta$ of the $\omega$ values in [2].

Uncertainties on the tabulated values: $\pm 0.2°$ on $\psi_{VD}$, $\psi_{VS}$; $\pm 0.007$ on $f_{VD}$, $f_{VS}$.

Table 3: Dimensionless functions $f_{VD}(\alpha,\beta,\eta)$, $f_{VS}(\alpha,\beta,\eta)$ and phase angles $\psi_{VD}$, $\omega$ and $\psi_{VS}$, for interface crack in symmetric sandwich with $\eta = h_1/h_c = 0.1$ and $\alpha = (\bar{E}_1 - \bar{E}_c)/(\bar{E}_1 + \bar{E}_c) \leq 0.8$ (additional results for $\alpha \geq 0.8$ in Tables 11-14)

| | \multicolumn{10}{c}{$\eta = h_1/h_c = 0.1$} |
|---|---|---|---|---|---|---|---|---|---|---|
| $\alpha$ | \multicolumn{2}{c|}{0} | \multicolumn{2}{c|}{0.2} | \multicolumn{2}{c|}{0.4} | \multicolumn{2}{c|}{0.6} | \multicolumn{2}{c|}{0.8} |
| $\beta$ | $f_{VD}$ | $f_{VS}$ | $f_{VD}$ | $f_{VS}$ | $f_{VD}$ | $f_{VS}$ | $f_{VD}$ | $f_{VS}$ | $f_{VD}$ | $f_{VS}$ |
| 0 | 1.926 | 1.882 | 2.064 | 2.007 | 2.264 | 2.191 | 2.591 | 2.499 | 3.282 | 3.165 |
| 0.1 | 1.955 | 1.908 | 2.093 | 2.031 | 2.293 | 2.213 | 2.619 | 2.516 | 3.308 | 3.172 |
| 0.2 | 1.980 | 1.930 | 2.118 | 2.052 | 2.317 | 2.231 | 2.642 | 2.529 | 3.328 | 3.172 |
| 0.3 | | | | | 2.338 | 2.245 | 2.661 | 2.538 | 3.343 | 3.168 |
| 0.4 | | | | | | | | | 3.353 | 3.158 |

| $\beta$ | $\psi_{VD}$ | $\omega$ | $\psi_{VS}$ | $\psi_{VD}$ | $\omega$ | $\psi_{VS}$ | $\psi_{VD}$ | $\omega$ | $\psi_{VS}$ | $\psi_{VD}$ | $\omega$ | $\psi_{VS}$ | $\psi_{VD}$ | $\omega$ | $\psi_{VS}$ |
|---|---|---|---|---|---|---|---|---|---|---|---|---|---|---|---|
| 0 | −0.2° | 52.2° | −1.1° | −1.5° | 54.2°* | −2.9° | −3.0° | 56.5°* | −5.0° | −4.5° | 60.2°* | −7.3° | −6.0° | 64.9°* | −10.2° |
| 0.1 | −2.6° | 47.8° | −3.5° | −3.8° | 49.9°* | −5.1° | −5.0° | 52.3°* | −6.9° | −6.1° | 56.3°* | −8.9° | −7.1° | 61.2°* | −11.3° |
| 0.2 | −5.1° | 43.2° | −6.0° | −6.1° | 45.5°* | −7.4° | −7.1° | 48.0°* | −8.9° | −7.9° | 52.2°* | −10.7 | −8.3° | 57.4°* | −12.5° |
| 0.3 | | | | | | | −9.3° | 43.5°* | −11.1 | −9.7° | 48.0°* | −12.4 | −9.6° | 53.5°* | −13.8° |
| 0.4 | | | | | | | | | | | | | −11.1 | 48.9°* | −15.2° |

*Interpolation/extrapolation on $\alpha,\beta$ of the $\omega$ values in [2].

Uncertainties on the tabulated values: ±0.2° on $\psi_{VD}$, $\psi_{VS}$; ±0.007 on $f_{VD}$, $f_{VS}$.

Table 4: Dimensionless functions $f_{VD}(\alpha,\beta,\eta)$, $f_{VS}(\alpha,\beta,\eta)$ and phase angles $\psi_{VD}$, $\omega$ and $\psi_{VS}$, for interface crack in symmetric sandwich with $\eta = h_1/h_c = 0.15$ and $\alpha = (\bar{E}_1 - \bar{E}_c)/(\bar{E}_1 + \bar{E}_c) \leq 0.8$ (additional results for $\alpha \geq 0.8$ in Tables 11-13)

| | \multicolumn{10}{c}{$\eta = h_1/h_c = 0.15$} |
|---|---|---|---|---|---|---|---|---|---|---|
| $\alpha$ | \multicolumn{2}{c|}{0} | \multicolumn{2}{c|}{0.2} | \multicolumn{2}{c|}{0.4} | \multicolumn{2}{c|}{0.6} | \multicolumn{2}{c|}{0.8} |
| $\beta$ | $f_{VD}$ | $f_{VS}$ | $f_{VD}$ | $f_{VS}$ | $f_{VD}$ | $f_{VS}$ | $f_{VD}$ | $f_{VS}$ | $f_{VD}$ | $f_{VS}$ |
| 0 | 1.927 | 1.849 | 2.066 | 1.970 | 2.268 | 2.153 | 2.596 | 2.461 | 3.290 | 3.131 |
| 0.1 | 1.956 | 1.872 | 2.094 | 1.991 | 2.296 | 2.171 | 2.625 | 2.473 | 3.317 | 3.130 |
| 0.2 | 1.980 | 1.892 | 2.119 | 2.009 | 2.321 | 2.184 | 2.648 | 2.480 | 3.338 | 3.123 |
| 0.3 | | | | | 2.341 | 2.194 | 2.667 | 2.483 | 3.353 | 3.110 |
| 0.4 | | | | | | | | | 3.363 | 3.091 |

| $\beta$ | $\psi_{VD}$ | $\omega$ | $\psi_{VS}$ | $\psi_{VD}$ | $\omega$ | $\psi_{VS}$ | $\psi_{VD}$ | $\omega$ | $\psi_{VS}$ | $\psi_{VD}$ | $\omega$ | $\psi_{VS}$ | $\psi_{VD}$ | $\omega$ | $\psi_{VS}$ |
|---|---|---|---|---|---|---|---|---|---|---|---|---|---|---|---|
| 0 | −0.4° | 52.1° | −2.0° | −1.8° | 54.2°* | −4.0° | −3.3° | 56.4°* | −6.3° | −4.9° | 60.0°* | −8.8° | −6.7° | 64.5°* | −12.0° |
| 0.1 | −2.8° | 47.8° | −4.4° | −4.0° | 49.8°* | −6.2° | −5.3° | 52.2°* | −8.2° | −6.6° | 56.1°* | −10.5 | −7.8° | 60.9°* | −13.1° |
| 0.2 | −5.4° | 43.2° | −6.9° | −6.4° | 45.4°* | −8.5° | −7.4° | 48.0°* | −10.3 | −8.3° | 52.1°* | −12.2 | −9.0° | 57.2°* | −14.4° |
| 0.3 | | | | | | | −9.6° | 43.5°* | −12.4 | −10.1 | 47.9°* | −13.9 | −10.2 | 53.3°* | −15.6° |
| 0.4 | | | | | | | | | | | | | −11.7 | 48.7°* | −17.0° |

*Values obtained through interpolation or extrapolation of the $\omega$ values in [2].

Uncertainties on the tabulated values: ±0.2° on $\psi_{VD}$, $\psi_{VS}$; ±0.007 on $f_{VD}$, $f_{VS}$.

Table 5: Dimensionless functions $f_{VD}(\alpha,\beta,\eta)$, $f_{VS}(\alpha,\beta,\eta)$ and phase angles $\psi_{VD}$, $\omega$ and $\psi_{VS}$, for interface crack in symmetric sandwich with $\eta = h_1/h_c = 0.25$ and $\alpha = (\bar{E}_1 - \bar{E}_c)/(\bar{E}_1 + \bar{E}_c) \leq 0.8$

| $\beta \backslash \alpha$ | 0 | | 0.2 | | 0.4 | | 0.6 | | 0.8 | |
|---|---|---|---|---|---|---|---|---|---|---|
|  | $f_{VD}$ | $f_{VS}$ | $f_{VD}$ | $f_{VS}$ | $f_{VD}$ | $f_{VS}$ | $f_{VD}$ | $f_{VS}$ | $f_{VD}$ | $f_{VS}$ |
| 0   | 1.937 | 1.788 | 2.078 | 1.907 | 2.280 | 2.089 | 2.608 | 2.397 | 3.294 | 3.071 |
| 0.1 | 1.967 | 1.806 | 2.107 | 1.922 | 2.310 | 2.100 | 2.639 | 2.402 | 3.328 | 3.060 |
| 0.2 | 1.992 | 1.821 | 2.133 | 1.934 | 2.336 | 2.107 | 2.665 | 2.401 | 3.354 | 3.043 |
| 0.3 |       |       |       |       | 2.357 | 2.109 | 2.685 | 2.395 | 3.373 | 3.018 |
| 0.4 |       |       |       |       |       |       |       |       | 3.385 | 2.987 |

| $\beta \backslash \alpha$ | $\psi_{VD}$ | $\omega$ | $\psi_{VS}$ | $\psi_{VD}$ | $\omega$ | $\psi_{VS}$ | $\psi_{VD}$ | $\omega$ | $\psi_{VS}$ | $\psi_{VD}$ | $\omega$ | $\psi_{VS}$ | $\psi_{VD}$ | $\omega$ | $\psi_{VS}$ |
|---|---|---|---|---|---|---|---|---|---|---|---|---|---|---|---|
| 0   | −0.7° | 52.0° | −3.5° | −2.2° | 54.0°* | −5.8° | −3.9° | 56.1°* | −8.3° | −5.8° | 59.4°* | −11.3 | −8.3° | 63.5°* | −15.0° |
| 0.1 | −3.1° | 47.6° | −5.9° | −4.4° | 49.7°* | −8.0° | −5.8° | 52.0°* | −10.2 | −7.3° | 55.6°* | −12.8 | −9.2° | 60.0°* | −16.1° |
| 0.2 | −5.6° | 43.1° | −8.4° | −6.7° | 45.3°* | −10.2° | −7.8° | 47.8°* | −12.2 | −9.0° | 51.7°* | −14.6 | −10.2 | 56.5°* | −17.3° |
| 0.3 |       |       |       |       |        |        | −10.0 | 43.4°* | −14.4 | −10.8 | 47.6°* | −16.3 | −11.4 | 52.7°* | −18.6° |
| 0.4 |       |       |       |       |        |        |       |        |       |       |        |       | −12.9 | 48.4°* | −20.0° |

\* Interpolation/extrapolation on $\alpha,\beta$ of the $\omega$ values in [2].

Uncertainties on the tabulated values: $\pm 0.2°$ on $\psi_{VD}$, $\psi_{VS}$; $\pm 0.007$ on $f_{VD}$, $f_{VS}$.

Table 6: Dimensionless functions $f_{VD}(\alpha,\beta,\eta)$, $f_{VS}(\alpha,\beta,\eta)$ and phase angles $\psi_{VD}$, $\omega$ and $\psi_{VS}$, for interface crack in symmetric sandwich with $\eta = h_1/h_c = 0.35$ and $\alpha = (\bar{E}_1 - \bar{E}_c)/(\bar{E}_1 + \bar{E}_c) \leq 0.8$

| $\beta \backslash \alpha$ | 0 | | 0.2 | | 0.4 | | 0.6 | | 0.8 | |
|---|---|---|---|---|---|---|---|---|---|---|
|  | $f_{VD}$ | $f_{VS}$ | $f_{VD}$ | $f_{VS}$ | $f_{VD}$ | $f_{VS}$ | $f_{VD}$ | $f_{VS}$ | $f_{VD}$ | $f_{VS}$ |
| 0   | 1.953 | 1.736 | 2.092 | 1.854 | 2.291 | 2.035 | 2.611 | 2.342 | 3.275 | 3.009 |
| 0.1 | 1.984 | 1.750 | 2.124 | 1.865 | 2.325 | 2.042 | 2.648 | 2.342 | 3.320 | 2.994 |
| 0.2 | 2.010 | 1.760 | 2.152 | 1.872 | 2.354 | 2.044 | 2.679 | 2.336 | 3.355 | 2.971 |
| 0.3 |       |       |       |       | 2.377 | 2.042 | 2.703 | 2.325 | 3.381 | 2.940 |
| 0.4 |       |       |       |       |       |       |       |       | 3.399 | 2.900 |

| $\beta \backslash \alpha$ | $\psi_{VD}$ | $\omega$ | $\psi_{VS}$ | $\psi_{VD}$ | $\omega$ | $\psi_{VS}$ | $\psi_{VD}$ | $\omega$ | $\psi_{VS}$ | $\psi_{VD}$ | $\omega$ | $\psi_{VS}$ | $\psi_{VD}$ | $\omega$ | $\psi_{VS}$ |
|---|---|---|---|---|---|---|---|---|---|---|---|---|---|---|---|
| 0   | −0.8° | 51.9° | −4.7° | −2.4° | 53.8°* | −7.2° | −4.4 | 55.8* | −10.0 | −6.8° | 58.7°* | −13.3 | −10.2 | 62.3°* | −17.7° |
| 0.1 | −3.2° | 47.5° | −7.1° | −4.6° | 49.5°* | −9.3° | −6.2 | 51.7* | −11.8 | −8.1° | 55.0°* | −14.8 | −10.8 | 59.0°* | −18.7° |
| 0.2 | −5.7° | 43.0° | −9.6° | −6.9° | 45.2°* | −11.6° | −8.2 | 47.6* | −13.9 | −9.7° | 51.2°* | −16.5 | −11.6 | 55.5°* | −19.9° |
| 0.3 |       |       |       |       |        |        | −10.3 | 43.2* | −15.9 | −11.4 | 47.1°* | −18.2 | −12.7 | 51.9°* | −21.1° |
| 0.4 |       |       |       |       |        |        |       |        |       |       |        |       | −14.0 | 47.7°* | −22.5° |

\* Interpolation/extrapolation on $\alpha,\beta$ and $\eta$ the $\omega$ values in [2].

Uncertainties on the tabulated values: $\pm 0.2°$ on $\psi_{VD}$, $\psi_{VS}$; $\pm 0.007$ on $f_{VD}$, $f_{VS}$.

Table 7: Dimensionless functions $f_{VD}(\alpha,\beta,\eta)$, $f_{VS}(\alpha,\beta,\eta)$ and phase angles $\psi_{VD}$, $\omega$ and $\psi_{VS}$, for interface crack in symmetric sandwich with $\eta = h_1/h_c = 0.5$ and $\alpha = (\bar{E}_1 - \bar{E}_c)/(\bar{E}_1 + \bar{E}_c) \leq 0.8$

| $\beta$ \ $\alpha$ | 0 | | 0.2 | | 0.4 | | 0.6 | | 0.8 | |
|---|---|---|---|---|---|---|---|---|---|---|
| | $f_{VD}$ | $f_{VS}$ | $f_{VD}$ | $f_{VS}$ | $f_{VD}$ | $f_{VS}$ | $f_{VD}$ | $f_{VS}$ | $f_{VD}$ | $f_{VS}$ |
| 0 | 1.978 | 1.671 | 2.112 | 1.789 | 2.301 | 1.966 | 2.602 | 2.265 | 3.217 | 2.914 |
| 0.1 | 2.013 | 1.681 | 2.149 | 1.796 | 2.342 | 1.970 | 2.650 | 2.262 | 3.280 | 2.898 |
| 0.2 | 2.042 | 1.686 | 2.181 | 1.798 | 2.377 | 1.968 | 2.690 | 2.252 | 3.331 | 2.871 |
| 0.3 | | | | | 2.405 | 1.960 | 2.722 | 2.236 | 3.371 | 2.840 |
| 0.4 | | | | | | | | | 3.401 | 2.788 |

| $\beta$ \ $\alpha$ | $\psi_{VD}$ | $\omega$ | $\psi_{VS}$ | $\psi_{VD}$ | $\omega$ | $\psi_{VS}$ | $\psi_{VD}$ | $\omega$ | $\psi_{VS}$ | $\psi_{VD}$ | $\omega$ | $\psi_{VS}$ | $\psi_{VD}$ | $\omega$ | $\psi_{VS}$ |
|---|---|---|---|---|---|---|---|---|---|---|---|---|---|---|---|
| 0 | −0.8° | 51.6° | −6.2° | −2.8° | 53.4°* | −8.9° | −5.2° | 55.3* | −12.0 | −8.3° | 57.7°* | −15.9 | −13.0 | 60.4°* | −21.6° |
| 0.1 | −3.1° | 47.3° | −8.5° | −4.8° | 49.2°* | −11.0° | −6.8° | 51.3* | −13.8 | −9.4° | 54.1°* | −17.3 | −13.1 | 57.4°* | −22.3° |
| 0.2 | −5.6° | 42.8° | −10.9° | −7.0° | 44.9°* | −13.2° | −8.7° | 47.2* | −18.8 | −10.7 | 50.4°* | −18.9 | −13.6 | 54.1°* | −23.3° |
| 0.3 | | | | | | | −10.7° | 42.9* | −17.7 | −12.3 | 46.4°* | −20.5 | −14.4 | 50.6°* | −24.4° |
| 0.4 | | | | | | | | | | | | | −15.6 | 46.7°* | −25.7° |

* Interpolation/extrapolation on $\alpha, \beta$ of the $\omega$ values in [2].

Uncertainties on the tabulated values: $\pm 0.2°$ on $\psi_{VD}$, $\psi_{VS}$; $\pm 0.007$ on $f_{VD}$, $f_{VS}$.

Table 8: Dimensionless functions $f_{VD}(\alpha,\beta,\eta)$, $f_{VS}(\alpha,\beta,\eta)$ and phase angles $\psi_{VD}$, $\omega$ and $\psi_{VS}$, for interface crack in symmetric sandwich with $\eta = h_1/h_c = 0.75$ and $\alpha = (\bar{E}_1 - \bar{E}_c)/(\bar{E}_1 + \bar{E}_c) \leq 0.8$

| $\beta$ \ $\alpha$ | 0 | | 0.2 | | 0.4 | | 0.6 | | 0.8 | |
|---|---|---|---|---|---|---|---|---|---|---|
| | $f_{VD}$ | $f_{VS}$ | $f_{VD}$ | $f_{VS}$ | $f_{VD}$ | $f_{VS}$ | $f_{VD}$ | $f_{VS}$ | $f_{VD}$ | $f_{VS}$ |
| 0 | 2.018 | 1.591 | 2.138 | 1.704 | 2.305 | 1.872 | 2.567 | 2.152 | 3.091 | 2.765 |
| 0.1 | 2.060 | 1.597 | 2.185 | 1.708 | 2.359 | 1.873 | 2.634 | 2.150 | 3.186 | 2.752 |
| 0.2 | 2.095 | 1.599 | 2.225 | 1.707 | 2.405 | 1.869 | 2.689 | 2.138 | 3.263 | 2.724 |
| 0.3 | | | | | 2.443 | 1.858 | 2.734 | 2.119 | 3.324 | 2.684 |
| 0.4 | | | | | | | | | 3.372 | 2.509 |

| $\beta$ \ $\alpha$ | $\psi_{VD}$ | $\omega$ | $\psi_{VS}$ | $\psi_{VD}$ | $\omega$ | $\psi_{VS}$ | $\psi_{VD}$ | $\omega$ | $\psi_{VS}$ | $\psi_{VD}$ | $\omega$ | $\psi_{VS}$ | $\psi_{VD}$ | $\omega$ | $\psi_{VS}$ |
|---|---|---|---|---|---|---|---|---|---|---|---|---|---|---|---|
| 0 | −0.7° | 51.3° | −7.8° | −3.2° | 52.8°* | −11.0° | −6.3° | 54.4* | −14.7 | −10.4 | 56.0°* | −19.5 | −16.5 | 57.5°* | −27.1° |
| 0.1 | −3.0° | 47.0° | −10.1° | −5.1° | 48.8°* | −12.9° | −7.7° | 50.5* | −16.3 | −11.0 | 52.6°* | −20.6 | −15.8 | 54.8°* | −27.4° |
| 0.2 | −5.4° | 42.5° | −12.4° | −7.2° | 44.5°* | −15.1° | −9.4° | 46.5* | −18.1 | −12.1 | 49.0°* | −22.1 | −15.9 | 51.7°* | −28.2° |
| 0.3 | | | | | | | −11.3° | 42.2* | −20.1 | −13.5 | 45.1°* | −23.6 | −16.4 | 48.4°* | −29.1° |
| 0.4 | | | | | | | | | | | | | −17.5 | 44.6°* | −33.9° |

* Interpolation/extrapolation on $\alpha, \beta$ and interpolation between $\eta = 0.5$ and $\eta = 1.0$ of the $\omega$ values [2].

Uncertainties on the tabulated values: $\pm 0.2°$ on $\psi_{VD}$, $\psi_{VS}$; $\pm 0.007$ on $f_{VD}$, $f_{VS}$.

Table 9: Dimensionless functions $f_{VD}(\alpha,\beta,\eta)$, $f_{VS}(\alpha,\beta,\eta)$ and phase angles $\psi_{VD}$, $\omega$ and $\psi_{VS}$, for interface crack in symmetric sandwich with $\eta = h_1/h_c = 1.0$ and $\alpha = (\bar{E}_1 - \bar{E}_c)/(\bar{E}_1 + \bar{E}_c) \leq 0.8$

| | \multicolumn{14}{c}{$\eta = h_1/h_c = 1.0$} |
|---|---|---|---|---|---|---|---|---|---|---|---|---|---|---|
| $\alpha$ \ $\beta$ | \multicolumn{2}{c}{0} | \multicolumn{2}{c}{0.2} | \multicolumn{2}{c}{0.4} | \multicolumn{2}{c}{0.6} | \multicolumn{2}{c}{0.8} | | | | | |
| | $f_{VD}$ | $f_{VS}$ | $f_{VD}$ | $f_{VS}$ | $f_{VD}$ | $f_{VS}$ | $f_{VD}$ | $f_{VS}$ | $f_{VD}$ | $f_{VS}$ | | | | |
| 0 | 2.052 | 1.533 | 2.157 | 1.640 | 2.301 | 1.797 | 2.527 | 2.059 | 2.975 | 2.637 | | | | |
| 0.1 | 2.101 | 1.539 | 2.212 | 1.644 | 2.367 | 1.799 | 2.609 | 2.058 | 3.096 | 2.628 | | | | |
| 0.2 | 2.142 | 1.539 | 2.260 | 1.642 | 2.422 | 1.794 | 2.678 | 2.047 | 3.193 | 2.601 | | | | |
| 0.3 | | | | | 2.468 | 1.782 | 2.734 | 2.027 | 3.270 | 2.561 | | | | |
| 0.4 | | | | | | | | | 3.330 | 2.509 | | | | |

| $\beta$ | $\psi_{VD}$ | $\omega$ | $\psi_{VS}$ | $\psi_{VD}$ | $\omega$ | $\psi_{VS}$ | $\psi_{VD}$ | $\omega$ | $\psi_{VS}$ | $\psi_{VD}$ | $\omega$ | $\psi_{VS}$ | $\psi_{VD}$ | $\omega$ | $\psi_{VS}$ |
|---|---|---|---|---|---|---|---|---|---|---|---|---|---|---|---|
| 0 | −0.7° | 51.6° | −9.0° | −3.5° | 53.4°* | −12.5° | −7.1° | 55.3* | −16.7 | −11.8 | 57.7°* | −22.4 | −18.3 | 60.4°* | −31.5° |
| 0.1 | −2.8° | 46.8° | −11.2° | −5.3° | 48.3°* | −14.3° | −8.3° | 49.8* | −18.2 | −12.1 | 51.1°* | −23.3 | −17.1 | 52.2°* | −31.7° |
| 0.2 | −5.3° | 42.8° | −13.5° | −7.3° | 44.9°* | −16.4° | −9.9° | 47.2* | −20.0 | −13.0 | 50.4°* | −24.6 | −17.0 | 54.1°* | −32.0° |
| 0.3 | | | | | | | −11.8° | 42.9* | −21.8 | −14.4 | 46.4°* | −26.1 | −17.4 | 50.6°* | −32.8° |
| 0.4 | | | | | | | | | | | | | −18.5 | 46.7°* | −33.9° |

\* Interpolation/extrapolation on $\alpha, \beta$ of the $\omega$ values in [2].

Uncertainties on the tabulated values: ±0.2° on $\psi_{VD}$, $\psi_{VS}$; ±0.007 on $f_{VD}$, $f_{VS}$.

Table 10: Composites with $\alpha \geq 0.8$. Dimensionless functions $f_{VD}(\alpha,\beta,\eta)$, $f_{VS}(\alpha,\beta,\eta)$ and phase angles $\psi_{VD}$, $\omega$ and $\psi_{VS}$, for interface crack in symmetric sandwich subjected to double shear load and $\eta = h_1/h_c = 0.01$.

| | \multicolumn{8}{c}{$\eta = h_1/h_c = 0.01$} |
|---|---|---|---|---|---|---|---|---|
| $\alpha$ \ $\beta$ | \multicolumn{2}{c}{0.8} | \multicolumn{2}{c}{0.9} | \multicolumn{2}{c}{0.99} | \multicolumn{2}{c}{0.998} |
| | $f_{VD}$ | $f_{VS}$ | $f_{VD}$ | $f_{VS}$ | $f_{VD}$ | $f_{VS}$ | $f_{VD}$ | $f_{VS}$ |
| 0.2 | 3.352 | 3.339 | 4.216 | 4.190 | 9.067 | 8.972 | 15.485 | 15.335 |
| 0.3 | 3.367 | 3.352 | 4.228 | 4.197 | 9.052 | 8.930 | 15.432 | 15.210 |
| 0.4 | 3.382 | 3.365 | 4.237 | 4.203 | 9.031 | 8.881 | 15.364 | 15.067 |

| $\beta$ | $\psi_{VD}$ | $\omega$ | $\psi_{VS}$ | $\psi_{VD}$ | $\omega$ | $\psi_{VS}$ | $\psi_{VD}$ | $\omega$ | $\psi_{VS}$ | $\psi_{VD}$ | $\omega$ | $\psi_{VS}$ |
|---|---|---|---|---|---|---|---|---|---|---|---|---|
| 0.2 | −6.5° | 57.8° | −6.9° | −5.9° | 63.2° | −6.7° | −2.2° | 76.3° | −4.9° | 0.3° | 81.9° | −3.7° |
| 0.3 | −7.7° | 53.8° | −8.1° | −6.6° | 59.3° | −7.4° | −1.3° | 74.5° | −4.0° | 2.3° | 80.7° | −1.7° |
| 0.4 | −9.2° | 49.2° | −9.6° | −7.5° | 55.7° | −8.3° | −0.4° | 72.1° | −3.0° | 4.4° | 79.5° | 0.4° |

\*Values obtained through interpolation or extrapolation of the $\omega$ values in [2].

Uncertainties on the tabulated values: ±0.2° on $\psi_{VD}$, $\psi_{VS}$; ±0.007 on $f_{VD}$, $f_{VS}$.

Table 11: Dimensionless functions $f_{VD}(\alpha,\beta,\eta)$, $f_{VS}(\alpha,\beta,\eta)$ and phase angles $\psi_{VD}$, $\omega$ and $\psi_{VS}$, for interface crack in symmetric sandwich with $\eta = h_1/h_c = 0.05$ and $\alpha = (\bar{E}_1 - \bar{E}_c)/(\bar{E}_1 + \bar{E}_c) \geq 0.8$

| | \multicolumn{11}{c}{$\eta = h_1/h_c = 0.05$} |
|---|---|---|---|---|---|---|---|---|---|---|---|
| $\alpha$ \ $\beta$ | \multicolumn{2}{c}{0.8} | | \multicolumn{2}{c}{0.9} | | \multicolumn{2}{c}{0.99} | | \multicolumn{2}{c}{0.998} | |
| | $f_{VD}$ | | $f_{VS}$ | $f_{VD}$ | | $f_{VS}$ | $f_{VD}$ | | $f_{VS}$ | $f_{VD}$ | $f_{VS}$ |
| 0.2 | 3.325 | | 3.240 | 4.193 | | 4.076 | 9.090 | | 8.885 | 15.664 | 15.517 |
| 0.3 | 3.341 | | 3.246 | 4.204 | | 4.069 | 9.066 | | 8.777 | 15.591 | 15.219 |
| 0.4 | 3.353 | | 3.248 | 4.209 | | 4.056 | 9.029 | | 8.653 | 15.488 | 14.886 |
| | $\psi_{VD}$ | $\omega$ | $\psi_{VS}$ | $\psi_{VD}$ | $\omega$ | $\psi_{VS}$ | $\psi_{VD}$ | $\omega$ | $\psi_{VS}$ | $\psi_{VD}$ | $\omega$ | $\psi_{VS}$ |
| 0.2 | −7.6° | 57.6°* | −10.1° | −7.2° | 62.7°* | −10.7° | −4.3° | 74.5°* | −10.9° | −3.3° | 78.8°* | −11.9° |
| 0.3 | −8.9° | 53.7°* | −11.3° | −7.9° | 59.0°* | −11.4° | −3.5° | 72.7°* | −10.1° | −1.4° | 78.0°* | −10.1° |
| 0.4 | −10.4° | 49.0°* | −12.7° | −8.8° | 55.4°* | −12.2° | −2.7° | 70.5°* | −9.2° | 0.5° | 77.2°* | −8.3° |

\* Interpolation/extrapolation on $\alpha,\beta$ and extrapolation on $\eta$ of the $\omega$ values in [2].

Uncertainties on the tabulated values: $\pm 0.2°$ on $\psi_{VD}$, $\psi_{VS}$; $\pm 0.007$ on $f_{VD}$, $f_{VS}$.

Table 12: Dimensionless functions $f_{VD}(\alpha,\beta,\eta)$, $f_{VS}(\alpha,\beta,\eta)$ and phase angles $\psi_{VD}$, $\omega$ and $\psi_{VS}$, for interface crack in symmetric sandwich with $\eta = h_1/h_c = 0.1$ and $\alpha = (\bar{E}_1 - \bar{E}_c)/(\bar{E}_1 + \bar{E}_c) \geq 0.8$ .

| | \multicolumn{11}{c}{$\eta = h_1/h_c = 0.1$} |
|---|---|---|---|---|---|---|---|---|---|---|---|
| $\alpha$ \ $\beta$ | \multicolumn{2}{c}{0.8} | | \multicolumn{2}{c}{0.9} | | \multicolumn{2}{c}{0.99} | | \multicolumn{2}{c}{0.998} | |
| | $f_{VD}$ | | $f_{VS}$ | $f_{VD}$ | | $f_{VS}$ | $f_{VD}$ | | $f_{VS}$ | $f_{VD}$ | $f_{VS}$ |
| 0.2 | 3.328 | | 3.172 | 4.204 | | 4.010 | 9.161 | | 8.929 | 15.617 | 15.994 |
| 0.3 | 3.343 | | 3.168 | 4.213 | | 3.987 | 9.139 | | 8.762 | 15.584 | 15.554 |
| 0.4 | 3.353 | | 3.158 | 4.216 | | 3.957 | 9.099 | | 8.575 | 15.509 | 15.063 |
| | $\psi_{VD}$ | $\omega$ | $\psi_{VS}$ | $\psi_{VD}$ | $\omega$ | $\psi_{VS}$ | $\psi_{VD}$ | $\omega$ | $\psi_{VS}$ | $\psi_{VD}$ | $\omega$ | $\psi_{VS}$ |
| 0.2 | −8.3° | 57.4°* | −12.5° | −8.1° | 62.2°** | −13.5° | −6.9° | 72.3°** | −15.7° | −8.9° | 73.8°** | −20.3° |
| 0.3 | −9.6° | 53.5°* | −13.8° | −8.9° | 58.6°** | −14.3° | −6.0° | 70.5°** | −15.0° | −6.7° | 73.2°** | −18.6° |
| 0.4 | −11.1° | 48.9°* | −15.2° | −9.8° | 55.0°** | −15.1° | −5.2° | 68.5°** | −14.3° | −4.5° | 72.6°** | −16.9° |

\* Interpolation/extrapolation on $\alpha,\beta$ of the $\omega$ values in [2]. \*\* Values in [2].

Uncertainties on the tabulated values: $\pm 0.2°$ on $\psi_{VD}$, $\psi_{VS}$; $\pm 0.007$ on $f_{VD}$, $f_{VS}$.

Table 13: Dimensionless functions $f_{VD}(\alpha,\beta,\eta)$, $f_{VS}(\alpha,\beta,\eta)$ and phase angles $\psi_{VD}$, $\omega$ and $\psi_{VS}$, for interface crack in symmetric sandwich with $\eta = h_1/h_c = 0.15$ and $\alpha = (\bar{E}_1 - \bar{E}_c)/(\bar{E}_1 + \bar{E}_c) \geq 0.8$

| | $\eta = h_1/h_c = 0.15$ | | | | | | | |
|---|---|---|---|---|---|---|---|---|
| $\alpha$ / $\beta$ | 0.8 | | 0.9 | | 0.99 | | 0.998 | |
| | $f_{VD}$ | $f_{VS}$ | $f_{VD}$ | $f_{VS}$ | $f_{VD}$ | $f_{VS}$ | $f_{VD}$ | $f_{VS}$ |
| 0.2 | 3.338 | 3.123 | 4.219 | 3.964 | 9.412 | 8.979 | 15.069 | 16.481 |
| 0.3 | 3.353 | 3.110 | 4.229 | 3.929 | 9.130 | 8.772 | 15.122 | 15.929 |
| 0.4 | 3.363 | 3.091 | 4.232 | 3.886 | 9.103 | 8.539 | 15.123 | 15.313 |

| | $\psi_{VD}$ | $\omega$ | $\psi_{VS}$ | $\psi_{VD}$ | $\omega$ | $\psi_{VS}$ | $\psi_{VD}$ | $\omega$ | $\psi_{VS}$ | $\psi_{VD}$ | $\omega$ | $\psi_{VS}$ |
|---|---|---|---|---|---|---|---|---|---|---|---|---|
| 0.2 | −9.0° | 57.2°** | −14.4° | −9.0° | 61.6°* | −15.7° | −10.0° | 69.9°* | −20.3° | −13.5° | 69.0°* | −28.9° |
| 0.3 | −10.2° | 53.3°** | −15.6° | −9.7° | 58.2°* | −16.5° | −8.9° | 68.2°* | −19.6° | −10.7° | 68.5°* | −27.3° |
| 0.4 | −11.7° | 48.7°** | −17.0° | −10.7° | 54.6°* | −17.4° | −8.0° | 66.4°* | −19.0° | −8.2° | 68.1°* | −25.6° |

\* Interpolation/extrapolation on $\alpha, \beta$ of the $\omega$ values in [2]. ** Values in [2].

Uncertainties on the tabulated values: ±0.2° on $\psi_{VD}$, $\psi_{VS}$; ±0.007 on $f_{VD}$, $f_{VS}$.

**Appendix A**

*A.1 Geometrical parameters, neutral axes and axial and bending stiffnesses*

The neutral axis of the composite substrate cross section, subscript "s", is calculated by imposing the absence of coupling between axial forces (bending moments) and bending curvature (axial strain). This can be done following the procedure in [6],[2] or simply imposing the bending-extensional coupling stiffness matrix, typically called **B**, of the laminate constitutive equations to be zero. The dimensional and dimensionless forms of the distance of the neutral axis from the centroid of the core are:

$$e_s = \frac{\bar{E}_1 h_1}{\bar{E}_c h_c + \bar{E}_1 h_1}\left(\frac{h_1}{2} + \frac{h_c}{2}\right) \quad \text{and} \quad \tilde{e}_s = \frac{e_s}{h_1} = \frac{\Sigma(\eta+1)}{2(\Sigma\eta+1)} \tag{35}$$

with $\Sigma$ and $\eta$ defined in Eq. (1).

The axial and bending stiffnesses of the different parts, delaminated "d", substrate "s" and base "b", calculated about the respective neutral axes are given below in dimensional and dimensionless forms:

$$A_d = \bar{E}_1 h_1 \quad \text{and} \quad \tilde{A}_d = A_d/(\bar{E}_1 h_1) = 1 \tag{36}$$
$$D_d = \frac{\bar{E}_1 h_1^3}{12} \quad \text{and} \quad \tilde{D}_d = D_d/(\bar{E}_1 h_1^3) = \frac{1}{12}$$

$$A_s = \bar{E}_1 h_1 + \bar{E}_c h_c \quad \text{and} \quad \tilde{A}_s = A_s / (\bar{E}_1 h_1) = 1 + \frac{1}{\eta \Sigma} \tag{37}$$

$$D_s = \bar{E}_1 h_1 \left[ \frac{h_1^2}{12} + \left( \frac{h_1}{2} + \frac{h_c}{2} - e_s \right)^2 \right] + \bar{E}_c h_c \left[ \frac{h_c^2}{12} + e_s^2 \right] \quad \text{and}$$

$$\tilde{D}_s = D_s / (\bar{E}_1 h_1^3) = \frac{1}{12} + \left( \frac{\eta+1}{2\eta} - \tilde{e}_s \right)^2 + \frac{1}{\Sigma \eta} \left( \frac{1}{12\eta^2} + \tilde{e}_s^2 \right)$$

$$A_b = \bar{E}_c h_c + 2\bar{E}_1 h_1 \quad \text{and} \quad \tilde{A}_b = A_b / (\bar{E}_1 h_1) = 2 + \frac{1}{\eta \Sigma} \tag{38}$$

$$D_b = 2\bar{E}_1 h_1 \left[ \frac{h_1^2}{12} + \left( \frac{h_1}{2} + \frac{h_c}{2} \right)^2 \right] + \frac{\bar{E}_c h_c^3}{12} \quad \text{and} \quad \tilde{D}_b = \frac{D_b}{\bar{E}_1 h_1^3} = 2 \left[ \frac{1}{12} + \frac{1}{\eta^2} \left( \frac{\eta}{2} + \frac{1}{2} \right)^2 \right] + \frac{1}{12 \Sigma \eta^3}$$

The coefficients which define the load sub-systems in Eqs. (2), given in dimensionless form in Eqs. (3), are:

$$M_* = M + P(h_1 / 2 + h_c / 2 + e_s) \tag{39}$$

$$C_1 = \frac{\bar{E}_1 h_1}{A_b} \tag{40}$$

$$C_2 = \frac{\bar{E}_1 h_1^2 (h_1 + h_c)}{2 D_b}$$

$$C_3 = \frac{\bar{E}_1 h_1^3}{12 D_b}$$

*A.2 Particularization to a homogeneous edge-cracked layer subjected to axial forces and bending moments:*

For homogeneous layer, assuming $\Sigma = 1$ in the equations above yield:

$$\tilde{A}_d = 1 \quad , \quad \tilde{D}_d = \frac{1}{12} \tag{41}$$

$$\tilde{e}_s = 0, \quad \tilde{A}_s = \frac{1}{\eta_H}, \quad \tilde{D}_S = \frac{1}{12\eta_H^3} \tag{42}$$

$$\tilde{A}_b = 1 + \frac{1}{\eta_H}, \quad \tilde{D}_b = \frac{1}{12} \left( 1 + \frac{1}{\eta_H} \right)^3 \tag{43}$$

$$M_* = M + P h_1 \left( \frac{1}{2} + \frac{1}{2\eta_H} \right)$$

$$C_1 = \frac{\eta_H}{1+\eta_H}, \quad C_2 = \frac{6}{\eta_H}\left(1+\frac{1}{\eta_H}\right)^{-3}, \quad C_3 = \left(1+\frac{1}{\eta_H}\right)^{-3} \tag{44}$$

where $\eta_H = h_1/(h_1+h_c)$ is the ratio of the thicknesses of the delaminated and substrate arm, which is related to $\eta$ through $\eta_H = \eta/(\eta+1)$. The dimensionless functions in the expression of the energy release rate are:

$$f_M = \sqrt{6(1+\eta_H^3)}, \quad f_P = \sqrt{\frac{1}{2}(1+4\eta_H+6\eta_H^2+3\eta_H^3)} \quad \text{and} \quad \sin\gamma_M = \frac{6\eta_H^2(\eta_H+1)}{f_M f_P} \tag{45}$$

And coincides with the solution in [11].

*A.3 Shear stiffnesses and shear correction factors (Jourawsky approximation)*

The shear stiffnesses of the debonded, substrate and delaminated arms are defined using the Jourawsky approximation as:

$$D_{Vd} = \kappa_{Vd} S_d \quad \text{and} \quad \tilde{D}_{Vd} = \frac{D_{Vd}}{\overline{E}_1 h} = \kappa_{Vd} \tilde{S}_d \tag{46}$$

$$D_{Vs} = \kappa_{Vs} S_s \quad \text{and} \quad \tilde{D}_{Vs} = \frac{D_{Vs}}{\overline{E}_1 h} = \kappa_{Vs} \tilde{S}_s$$

$$D_{Vb} = \kappa_{Vb} S_b \quad \text{and} \quad \tilde{D}_{Vb} = \frac{D_{Vb}}{\overline{E}_1 h} = \kappa_{Vb} \tilde{S}_b$$

with

$$S_1 = S_d = h_1 G_1 \quad \text{and} \quad \tilde{S}_d = \frac{S_d}{\overline{E}_1 h} = \frac{G_1}{\overline{E}_1} = \frac{1-v_1}{2} \tag{47}$$

$$S_2 = S_s = h_1 G_1 + h_c G_c \quad \text{and} \quad \tilde{S}_s = \frac{S_s}{\overline{E}_1 h} = \frac{1-v_1}{2} + \frac{1}{2\Sigma\eta}(1-v_c)$$

$$S_3 = S_b = 2h_1 G_1 + h_c G_c \quad \text{and} \quad \tilde{S}_b = \frac{S_b}{\overline{E}_1 h} = 1-v_1 + \frac{1}{2\Sigma\eta}(1-v_c)$$

The shear correction factors of the different arms are calculated by imposing the equality of the elastic energies associated to the shear forces and calculated using the global and local variables:

$$\frac{1}{2}\gamma_i V_i = \frac{1}{2}\int_h 2\sigma_{xy}(\zeta_i)\varepsilon_{xy}(\zeta_i)d\zeta_i \tag{48}$$

with $i=1=d$, $i=2=s$, $i=3=b$ and $\gamma_i = V_i / D_{Vi}$ the global shear strain; the coordinate $\zeta_i$ is a vertical upward coordinate with origin in the neutral axis of the arm $i$ and the integral is calculated over the whole height of each arm.

The Jourawsky approximation is used for the calculation of the shear stresses in Eq.(48) from the exact bending stresses, $\sigma_{xx}(\zeta_i) = \dfrac{M_i}{D_i}\zeta_i \bar{E}_i + \dfrac{N_i}{A_i}\bar{E}_i$, through the imposition of local equilibrium, $\sigma_{xy,y} + \sigma_{xx,x} = 0$ and knowing that $\dfrac{\partial M_i}{\partial x} = V_i$ and $\dfrac{\partial N_i}{\partial x} = 0$ from global equilibrium. Some algebraic manipulations yields $\kappa_{Vd} = \kappa_{V1} = 5/6$ for the debonded homogeneous layer and the following formulas for the substrate and bonded layers, i=2,3:

$$\kappa_{Vi} = \left[ \dfrac{S_i}{D_{Vi}^2} \left( \int_{-d_i}^{-d_i+h_1} \dfrac{l(\zeta_i)^2}{G_1} d\zeta_i + \int_{-d_i+h_1}^{-d_i+h_1+h_c} \dfrac{m(\zeta_i)^2}{G_c} d\zeta_i + \int_{-d_i+h_1+h_c}^{-d_i+h_c+2h_1} \dfrac{n_i(\zeta_i)^2}{G_1} d\zeta_i \right) \right]^{-1} \quad (49)$$

$$l(\zeta_i) = -\dfrac{\bar{E}_1}{2}\zeta_i^2 + \dfrac{\bar{E}_1 d_i^2}{2}$$

$$m(\zeta_i) = -\dfrac{\bar{E}_c}{2}\zeta_i^2 + \dfrac{\bar{E}_1 d_i^2}{2} + \dfrac{\bar{E}_c - \bar{E}_1}{2}(h_1 - d_i)^2$$

$$n_i(\zeta_i) = \begin{cases} 0 \quad \text{for } i=2, \text{ in substrate} \\ -\dfrac{\bar{E}_1}{2}\zeta_i^2 + \dfrac{\bar{E}_1 d^2}{2} + \dfrac{\bar{E}_c - \bar{E}_1}{2}(h_1 - d_i)^2 + \dfrac{\bar{E}_1 - \bar{E}_c}{2}(h_1 + h_c - d_i)^2 \quad \text{for } i=3, \text{ in base layer} \end{cases}$$

where $d_3 = \dfrac{h_c}{2} + h_1$ and $d_2 = \dfrac{h_c}{2} + h_1 - e_s$ are the distances of the neutral axes of the bonded and substrate parts from the intradox, and $e_s$ is the distance of the neutral axis of the substrate section from the centroid of the core, Eq. (35) and Fig. 2.

The normalized shear stiffnesses and the shear correction factors for the problem examined in Section 5, with $\eta = h_1/h_c = 0.25$, $\Sigma = \bar{E}_1/\bar{E}_c = 9$ (or $\alpha = 0.8$) and $\beta = 0$ depend on the Poisson ratios of the layers. For $\nu_1 = \nu_c = 0.5$ (incompressible layers), Eq. (49) yields $\kappa_{Vd} = 5/6$, $\kappa_{Vs} = 0.376$ and $\kappa_{Vb} = 0.28$; the corresponding dimensionless shear stiffnesses are $\tilde{D}_{Vd} = 5/6 \tilde{S}_d = 0.208$, $\tilde{D}_{Vs} = \kappa_{Vs}\tilde{S}_s = 0.136$ and $\tilde{D}_{Vb} = \kappa_{Vb}\tilde{S}_b = 0.171$. The contribution of the shear strains in Eq. (34) is then $\dfrac{1}{2}(1/\tilde{D}_{Vd} + 1/\tilde{D}_{Vs}) = 6.08$, which yields $(a_1^{VD} - a_2^{VD}) = 4.77$ since $f_{VD}^2 = 10.85$, Eq. (30). For $\nu_1 = 0.2$ and $\nu_c = 0.478$, Eq. (49) yields $\kappa_{Vd} = 5/6$, $\kappa_{Vs} = 0.277$ and $\kappa_{Vb} = 0.196$ the dimensionless stiffnesses $\tilde{D}_{Vd} = \tilde{S}_d 5/6 = 0.333$, $\tilde{D}_{Vs} = \tilde{S}_s \kappa_{Vs} = 0.143$ and $\tilde{D}_{Vb} = \tilde{S}_b \kappa_{Vb} = 0.18$. The contribution of the shear strains in Eq. (34) is then $\frac{1}{2}\left(1/\tilde{D}_{Vd} + 1/\tilde{D}_{Vs}\right) = 5.$, which yields $\left(a_1^{VD} - a_2^{VD}\right) = 5.85$.

**A.4 *Verification of the accuracy of proposed methodology and tabulated results***

The coefficients $f_{VD}, f_{VS}, \psi_{VD}, \psi_{VS}$ derived using the model presented in Section 6.1 for a bi-material beam with $\alpha = -0.8 \div 0.8$, $\beta = 0$ and $\eta_{bi} = 0 \div 1$, with $\eta_{bi} = h_1/(h_1 + h_c) = \eta/(\eta+1)$, have been compared with those derived in [7] through ABAQUS and the Virtual Crack Extension method [22][23]. The coefficients were found in agreements within the uncertainties defined in the Tables 2,3 in [7], namely ±0.007 on $f_{VD}, f_{VS}$ and ±0.2° on $\psi_{VD}, \psi_{VS}$. Similar agreement has been found in the prediction of the angle $\omega$ for a bi-material beam with $\alpha = -0.8 \div 0.8$, $\beta = 0$ and $\eta_{bi} = 0 \div 1$ from FE analyses of beams subjected to pure bending moments M, Table 1 in [7].

The coefficients in the Tables 1-9 have also been successfully compared with the results presented in Table 1 and Eqs. 6 and 14 in [8] and obtained from root rotations coefficients calculated numerically using ANSYS for a homogeneous isotropic edge cracked layer with $\alpha = 0$, $\beta = 0$ and $\eta_H = 0 \div 1$, with $\eta_H = h_1/(h_1 + h_c)$.

For symmetric sandwich beams, the results on the angle $\omega$ presented in [2] on varying $\alpha = 0.8 \div 0.998$, $\beta = 0.2$ and $\eta = 0.1$ and $1.0$, obtained using ANSYS and the CSDE method, have been successfully reproduced through direct comparison for all cases with common parameters and linear interpolation for the other cases. Other spot verifications have been conducted for different values of $\alpha, \beta, \eta$.

The accuracy on the values of the phase angle $\omega$ obtained through interpolation/extrapolation of the results in [2] have been verified with accurate finite element analyses. For instance, the value presented in Table 1 for $\eta = 0.025$, $\alpha = 0.8$ and $\beta = 0.4$ has been obtained through a linear extrapolation from the values for $\eta = 0.1$ and $\eta = 0.125$ in [2] and virtually coincides with the value obtained using the model in Section 6.

In addition, the accuracy of the methodology formulated in this paper to calculate fracture parameters in sandwich beams subjected to arbitrary loading conditions and to analyze typical laboratory fracture specimen, has been verified by comparing the predictions of Eqs. (20),(21) with the FE results obtained using ANSYS and the CSDE method in [3] for DCB-UBM bi-material and sandwich specimens subjected to uneven bending moments. The relative differences on G were found to be always below 2% on varying the ratio of the applied moments for aluminum face sheet/aluminum core bi-material and sandwich specimens, and below 4% for aluminum face-sheets/H100 foam core sandwich beams. Additional verifications on the accuracy of the method proposed in this paper, using FE analyses on DCB sandwich specimens, have been presented in Section 5.